\documentclass[12pt]{iopart}
\usepackage{iopams}
\usepackage{graphicx,hyperref,cite,soul}
\usepackage[dvipsnames]{xcolor}
\usepackage[titletoc,title]{appendix} 
\usepackage[normalem]{ulem}

\newcommand{\T}{\mathcal{T}}
\newcommand{\eqref}[1]{\eref{#1}}
\newcommand{\expval}[1]{\langle #1 \rangle}
\newcommand{\dd}[1]{{\mathrm{d} #1}}
\newcommand{\dv}[3]{\frac{\mathrm{d}^{#1} #2}{\mathrm{d} {#3}^{#1}}}

\begin{document}

\title[Short title]{First-Passage-Time Asymmetry for Biased Run-and-Tumble Processes}
\author{Yonathan Sarmiento$^{1,2}$, Benjamin Walter$^{3}$, Debraj Das$^{1,2}$\footnote{Corresponding author: \mailto{ddas@sissa.it}}, Samvit Mahapatra$^4$, \'Edgar Rold\'an$^1$ and Rosemary J. Harris$^5$}
\address{$^1$ ICTP -- The Abdus Salam International Centre for Theoretical Physics, Strada Costiera 11, 34151 Trieste, Italy}
\address{$^2$ International School for Advanced Studies (SISSA), Via Bonomea 265, 34136 Trieste, Italy}
\address{$^3$ Department of Mathematics, Imperial College London, London SW7 2AZ, United Kingdom}
\address{$^4$ Department of Physics, Ravenshaw University, Cuttack, Odisha 753003, India}
\address{$^5$ Department of Mathematics, University College London, Gower Street, London WC1E~6BT, United Kingdom}
\eads{\mailto{ddas@sissa.it}, \mailto{edgar@ictp.it}, \mailto{rosemary.j.harris@ucl.ac.uk}}
\vspace{10pt}
\begin{indented}
\item[]\today
\end{indented}

\begin{abstract}
We explore first-passage phenomenology for biased active processes with a renewal-type structure, focusing in particular on paradigmatic run-and-tumble models in both discrete and continuous state spaces. In general, we show there is no equality between distributions of conditional first-passage times to symmetric barriers positioned in and against the bias direction. However, we give conditions for such a duality to be restored asymptotically (in the limit of a large barrier distance) and highlight connections to the Gallavotti-Cohen fluctuation relation and the method of images. Our general trajectory arguments  of first-passage-time distributions for asymmetric run-and-tumble processes to escape from an interval of arbitrary width are supported by exact analytical results, which we derive extending Montroll's defect technique. Furthermore, we quantify the degree of violation of first-passage duality using Kullback-Leibler divergence and signal-to-noise ratios associated with the first-passage times to the two barriers. We reveal an intriguing dependence of such measures of first-passage asymmetry on the underlying often hidden tumbling dynamics which may inspire inference techniques based on first-passage-time statistics in active systems. 

\end{abstract}

\newpage
%\maketitle
\tableofcontents

%================================================
%================================================
%================================================
\section{Introduction}
\label{s:intro}

Symmetries ranging from the trivial to the surprising have long played an important role in theoretical physics.  In the particular context of non-equilibrium stochastic processes, one symmetry which has attracted recent attention is the so-called `first-passage duality', `first-passage-time fluctuation theorem' or `generalized Haldane equality'~\cite{qian2006generalized,ge2012stochastic,bauer2014affinity,neri2017statistics,krapivsky2018first}. This family of results involves statistics associated with different variants of first-passage times. In one dimension, consider for example a biased random walker starting at the origin, with absorbing boundaries at $\pm L$; the symmetry states that the distribution of first-passage times to $+L$ (given the particle reaches that upper boundary first) is \emph{identical} to the distribution of first-passage times to $-L$ (given the particle reaches the lower boundary first). Such symmetry implies that it takes the same amount  of time on average to first reach the boundary in the direction of the bias than to first reach the boundary in the direction opposite to the bias (and similarly for any moment of the first-passage time). At first sight this symmetry perhaps falls in the surprising category although, as we explain later, it can actually be understood quite straightforwardly for simple random walks. Applications to higher-dimensional random walks~\cite{johnson2023first} as well as (integrated) current-like quantities in more general Markov processes are also readily available~\cite{harunari2022learn}.  In the following we will abbreviate `first-passage time' as FPT and generically refer to the symmetry of interest as \emph{FPT duality} (to be defined more precisely in the next section).

Another area of great topical interest in statistical mechanics is the study of reset processes, see for example the recent reviews~\cite{Evans_2020, Nagar_2023}.  There is a considerable body of work focused on particles resetting their position to a fixed location (or distribution of locations) at stochastic intervals as relevant for some search processes.  However, the mathematical machinery of renewal processes (in which Laplace transforms are key) can be analogously applied to other forms of reset.  Particularly significant are run-and-tumble processes which serve as paradigms for modelling the motion of certain kinds of bacteria~\cite{Berg_2004,Cates_2012}.  The idea here is that a bacterium moves, possibly noisily, in some preferred direction (a `run') before pausing and using its flagella to perform a stochastic reorientation of that preferred direction (a `tumble').  More generally, we can think of a class of models consisting of a Markov process with preferred direction, interrupted by resets of direction.\footnote{Even more generally, we could widen the class to allow resets of velocity, as in~\cite{Mori_2021}.}  If the resets occur with constant probability in discrete time (or constant rate in continuous time) then such processes are, of course, Markovian in the extended state space of position and preferred direction but if we consider only the position state space (corresponding perhaps to experimental observation of a spatial trajectory) then they are non-Markovian with the preferred direction as a `hidden' state.  One can also extend the framework to run-and-tumble processes with non-constant tumble probabilities/rates (i.e., non-geometrically/non-exponentially distributed run lengths) which are non-Markovian even when the preferred direction is included in the state space.  

We further note that run-and-tumble particles are examples of `active' matter in the sense that they are not just driven passively by some external field but can change their own motion through an internal process (physically requiring some energy source). The application of statistical mechanics to active processes is also an important current topic, particularly in the context of biological modelling~\cite{Ramaswamy_2010,Vrugt_2024}.  

In this contribution we aim to bring the subjects of the preceding paragraphs together and examine the validity of the FPT duality in reset processes. To be concrete, we focus here on one-dimensional single-particle run-and-tumble models although much of the analysis could presumably also be extended to current-like quantities in other renewal processes. In fact, we show that the renewal structure allows a general prediction about the conditions under which the duality holds asymptotically in the large-$L$ (wide-interval) limit; essentially this follows directly from a correspondence with the Gallavotti-Cohen fluctuation relation~\cite{Lebowitz99} as already pointed out for Markov processes in the seminal work of Gingrich and Horowitz~\cite{gingrich2017fundamental}. Our simple models allow us to demonstrate how the argument applies in different cases with/without asymptotic duality and also to illuminate finite-size corrections which may be more relevant practically -- here measuring asymmetry could provide hints about the nature of the underlying dynamics. Some related recent discussion about the experimental relevance of FPT-duality breaking, and the form of corrections, can be found in~\cite{piephoff2025} in the context of biomolecular networks. 

The structure of the rest of the paper is as follows. In section~\ref{s:setup} we define some notation and explain more precisely the symmetry property of interest as well as introducing simple random walk and run-and-tumble models which will serve as examples in the remainder of the paper; we also present various alternative ways to quantify first-passage asymmetry, even when full statistics are not available. In section~\ref{s:symmetry} we discuss how the FPT duality can be understood by considering trajectory symmetries under various transformations, highlighting along the way connections to both the Gallavotti-Cohen fluctuation relation and the method of images; by applying these arguments specifically to renewal processes, we then obtain conditions for the asymptotic validity of the duality. Significantly, in section~\ref{s:fpt-theory}, we obtain exact (albeit sometimes cumbersome) analytical expressions for the first-passage statistics of a discrete run-and-tumble model with bias; we then compare these to numerics in section~\ref{s:results} showing convincing agreement for the chosen measures of asymmetry. In section~\ref{s:scaling} we focus on the scaling behaviour of these results and discuss bounds for the asymptotic behaviour. In section~\ref{s:conc}, we outline predictions for various modified run-and-tumble models, provide a concluding perspective, and suggest some open questions. Finally, the appendices contain more details of the renewal argument as well as some additional calculations for run-and-tumble models, in both discrete and continuous time.

Note that during the final stages of preparation of this manuscript a closely-related contribution by Neri appeared in the literature~\cite{Neri_2025}. That work uses a `Perron martingale' approach to develop an elegant general argument for first-passage properties of continuous-time processes, with one of the explicit examples being a symmetric run-and-tumble model with asymmetrically-positioned absorbing boundaries. Our work is complementary in the sense that we consider mainly discrete-time processes and look at asymmetric run-and-tumble models with symmetric boundaries. We also refer readers to the recent work of two of us~\cite{Lea2025} where analytical formulae for various FPT statistics associated with discrete run-and-tumble models reaching a single absorbing boundary were obtained in the context of quantum thermodynamics.

%================================================
%================================================
%================================================
\section{Framework}
\label{s:setup}

We here introduce the key ingredients and ideas which play important roles in this paper.

\subsection{Notation and quantities of interest}
\label{ss:quantities}
\begin{figure}
\centering
\includegraphics[scale=0.55]{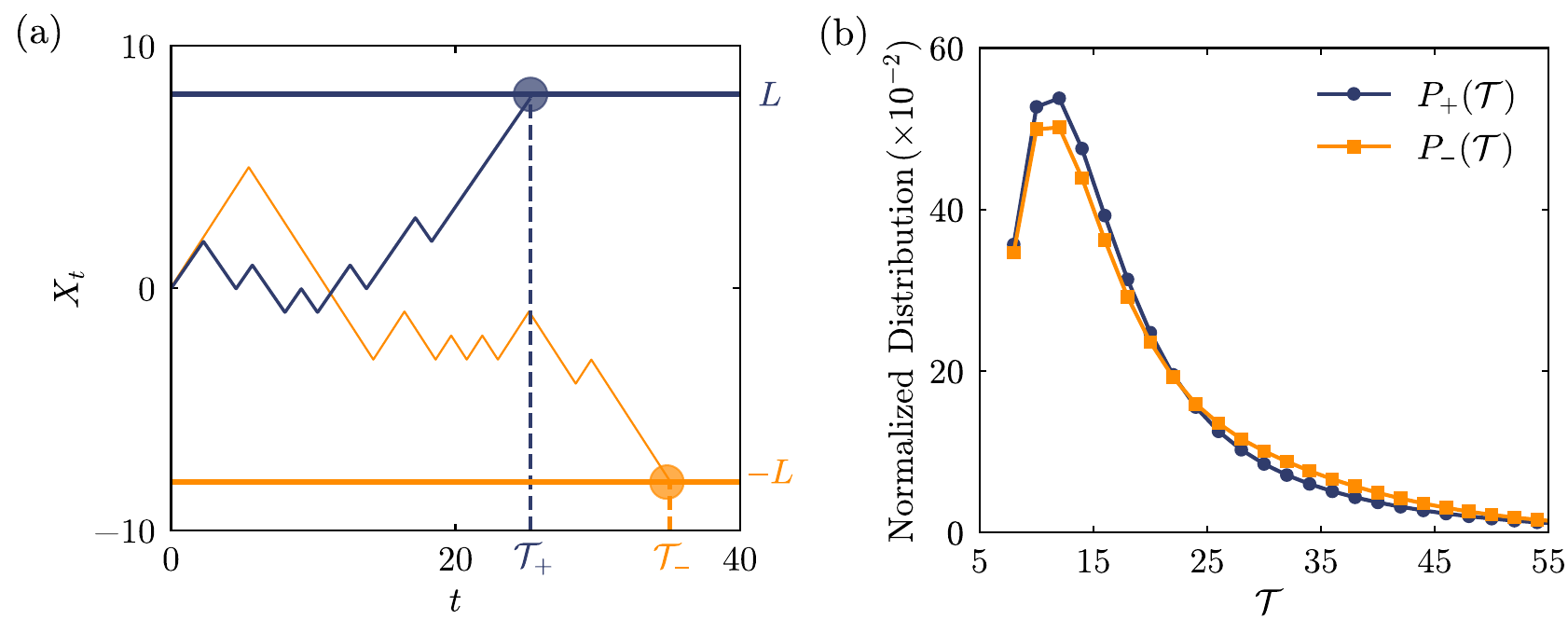}
\caption{FPT illustration for the position of a single particle on the symmetric interval~$[-L, L]$. As a representative example, we use the discrete run-and-tumble (RnT) model defined in section~\ref{ss:RnT}, with parameters: $L=8$, $p=0.7$, $p'=0.75$, and $f=0.05$. (a) The process stops at a stochastic time $T$ given by the earliest time when the particle reaches either of two absorbing boundaries, $+L$ or $-L$. The solid zig-zag  blue and orange lines are sample trajectories escaping the interval via the boundaries $+L$  and $-L$ at times $\T_+$  and $\T_-$, respectively. (b) Normalized  distributions $P_+(\T)$ (blue filled circles) and $P_-(\T)$ (orange filled squares), associated with  $\T_+$  and $\T_-$ respectively are displayed  with symbols corresponding to simulations from $10^8$ trajectories. The lines are a guide to the eye which join discrete points given by the exact analytical expressions obtained from numerically inverting the $z$-transforms of $P_+(\T)$ and $P_-(\T)$~\eqref{eq:F-z}, see section~\ref{s:fpt-theory} for further details.  Here, $\langle T \rangle_+\approx 19$, $\langle T \rangle_- \approx 20$, $\mathrm{Var}_+(T) \approx 134$, $\mathrm{Var}_-(T) \approx 156$, which indicates that the FPT duality given by~\eqref{e:duality} does not hold in this example.}
\label{f:traj}
\end{figure}

Throughout this work we consider a single particle in one spatial dimension with position at time $t$ given by the random variable $X_t$. We generally start the particle at the origin (so $X_0=0$) and place absorbing boundaries at $+ L$ and $-L$. In the main text we focus on processes in discrete time and space such that both $t$ and $X$ take integer values. Representative trajectories for a process in this class are shown in figure~\ref{f:traj}(a).

We are interested in the time at which the particle exits the interval $(-L,L)$ by being absorbed at either boundary. We denote this first-passage (or first-exit) time by the random variable $T$ with probability mass function  $P(\mathcal{T})=\mathrm{Prob}(T=\mathcal{T})$. Using angular brackets to denote averages over histories, we define scaled moments $\bar{\tau}$ and $\sigma^2$ such that 
\begin{equation}
\langle T \rangle = \bar{\tau} L, \qquad 
\mathrm{Var}(T)= \langle T^2 \rangle - \langle T \rangle^2 = \sigma^2 L.
\end{equation}
Note that, by construction, we have $X_{T}= + L$ when the particle is absorbed at the upper boundary and  $X_{T}= - L$  when the particle is absorbed at the lower boundary; it is also convenient to have a scaled indicator variable $B=X_{T}/L$ which takes the values $\pm1$. For ease of notation throughout the text, we will often use $\pm$ as shorthand labelling for $\pm 1$, and loosely write equations like `$B=+$'. The respective splitting probabilities are then $\mathsf{P}_+ = \mathrm{Prob}(B=+)$ and $\mathsf{P}_-=\mathrm{Prob}(B=-)$. We usually take positive bias such that $\mathsf{P}_+ > \mathsf{P}_-$. 

To consider the first-passage symmetry, the key objects are the \emph{conditional} distributions at the boundaries which we write as
\begin{equation}
P_\pm(\mathcal{T})=\mathrm{Prob}(T=\mathcal{T}|B=\pm). \label{e:fpt_splitting_def}
\end{equation} 
We also use $+$ and $-$ subscripts to denote expectations and moments with respect to these conditional distributions.  The basic statement of the FPT duality is simply that, 
\begin{equation}
P_+(\mathcal{T})=P_-(\mathcal{T})
    \label{e:duality}
\end{equation}
which we can write explicitly as  $\mathrm{Prob}(T=\mathcal{T}|B=+)=\mathrm{Prob}(T=\mathcal{T}|B=-),$ or, equivalently,
\begin{equation}
\frac{\mathrm{Prob}(T=\mathcal{T},B=+)}{\mathrm{Prob}(B=+)}=\frac{\mathrm{Prob}(T=\mathcal{T},B=-)}{\mathrm{Prob}(B=-)}.
\end{equation}
We emphasize again that the equality~\eqref{e:duality} is at the level of the conditional distributions; for a biased process it is obviously generally not true that the unconditioned probability of being absorbed at $+L$ at time $\mathcal{T}$ is the same as the probability of being absorbed at $-L$ at time $\mathcal{T}$, i.e., in our notation $\mathrm{Prob}(T=\mathcal{T},B=+)$ is generally not equal to $\mathrm{Prob}(T=\mathcal{T},B=-)$.  

To fix ideas, one can imagine generating a large ensemble of trajectories for some process via computer simulation and building empirical histograms of the absorption times at upper and lower boundaries; with a positive bias there will be more data points for the upper histogram but the claim is that if FPT duality holds the \emph{normalized} histograms (cf.\ those shown in figure~\ref{f:traj}(b) above) should be identical.  This means, of course, that the corresponding conditioned moments should also be the same; in particular, if duality holds we have  $\langle T \rangle_+=\langle T \rangle_-$ and $\mathrm{Var}_+(T)=\mathrm{Var}_-(T)$ (with obvious equivalents for the scaled moments).  We clearly see that duality fails for the particular distributions in figure~\ref{f:traj}(b); in the remainder of the paper we seek to explore this symmetry-breaking more generally.

\subsection{Example models}
\label{ss:models}

Before examining the conditions for the FPT duality to hold/fail, we first define two concrete models to serve as minimal yet enlightening examples.

\subsubsection{RW: minimal random-walk model}
\label{sss:RW}
For comparison purposes, it is helpful to have in mind perhaps the most basic  non-equilibrium model: a simple biased random walk in one dimension, in which the particle jumps at every time step in the positive direction (`up' when we plot trajectories as in figure~\ref{f:traj} but sometimes also called `right' in view of the usual $x$-axis orientation) with probability $p$ and the negative direction (`down'/`left') with probability $q=1-p$.  The transition probabilities (conditional probability mass functions) are thus given by
\begin{eqnarray}
    P(x+1|x) &= \mathrm{Prob}(X_{t+1}=x+1| X_t=x) = p , \\
    P(x-1|x) &= \mathrm{Prob}(X_{t+1}=x-1| X_t=x) =q. 
\end{eqnarray}
Of course, this process (hereafter referred to as the \emph{RW model}) is already rather well-understood but we shall use it later to demonstrate heuristically how the FPT duality emerges from trajectory properties. For now, we merely note that when $p>q$ there is a strictly positive bias such that the particle is more likely to be absorbed at the upper boundary than the lower boundary.  

\subsubsection{RnT: minimal run-and-tumble model}
\label{ss:RnT}

We now build a discrete run-and-tumble  model by considering random walk dynamics as above, with probability $p'$ ($q'=1-p'$) for moving in (against) the preferred direction and the added twist that at the beginning of every time step there is the possibility for this preferred direction to change.  Specifically, at every time step the process `tumbles' with probability $f$ and does not tumble with probability $1-f$.  If a tumble occurs the preferred direction (which can be thought of as a kind of `spin' variable) is reset: positive/right with probability $p$ and negative/left with probability $q=1-p$.  Henceforth we refer to this discrete run-and-tumble process simply as the \emph{RnT model}; it is a variant of the model introduced in~\cite{shreshtha2019thermodynamic}.

Of course, in this one-dimensional discrete setting (in contrast to run-and-tumble models in higher dimension and continuous space~\cite{Sevilla_2020,Santra_2020}), it is quite likely that a `tumble' does not actually change the preferred direction (since there is no restriction on the reset dynamics choosing the same direction as in the previous tumble) and we will later find it useful to consider `effective tumbles' as those which actually do change the preferred direction, from negative to positive with probability $fp$, or positive to negative with probability $fq$.  We note that the tumbling is instantaneous in the sense that, even if the preferred direction is switched in a particular time step, a random-walk move still takes place in the same time step (with the new preferred direction); an arguably more realistic model in which the tumble operation causes the particle to remain stationary for one time step (latent time) was discussed in~\cite{shreshtha2019thermodynamic} and will be mentioned again in section~\ref{s:conc}. 

\begin{figure}
\centering
\includegraphics[scale=0.35]{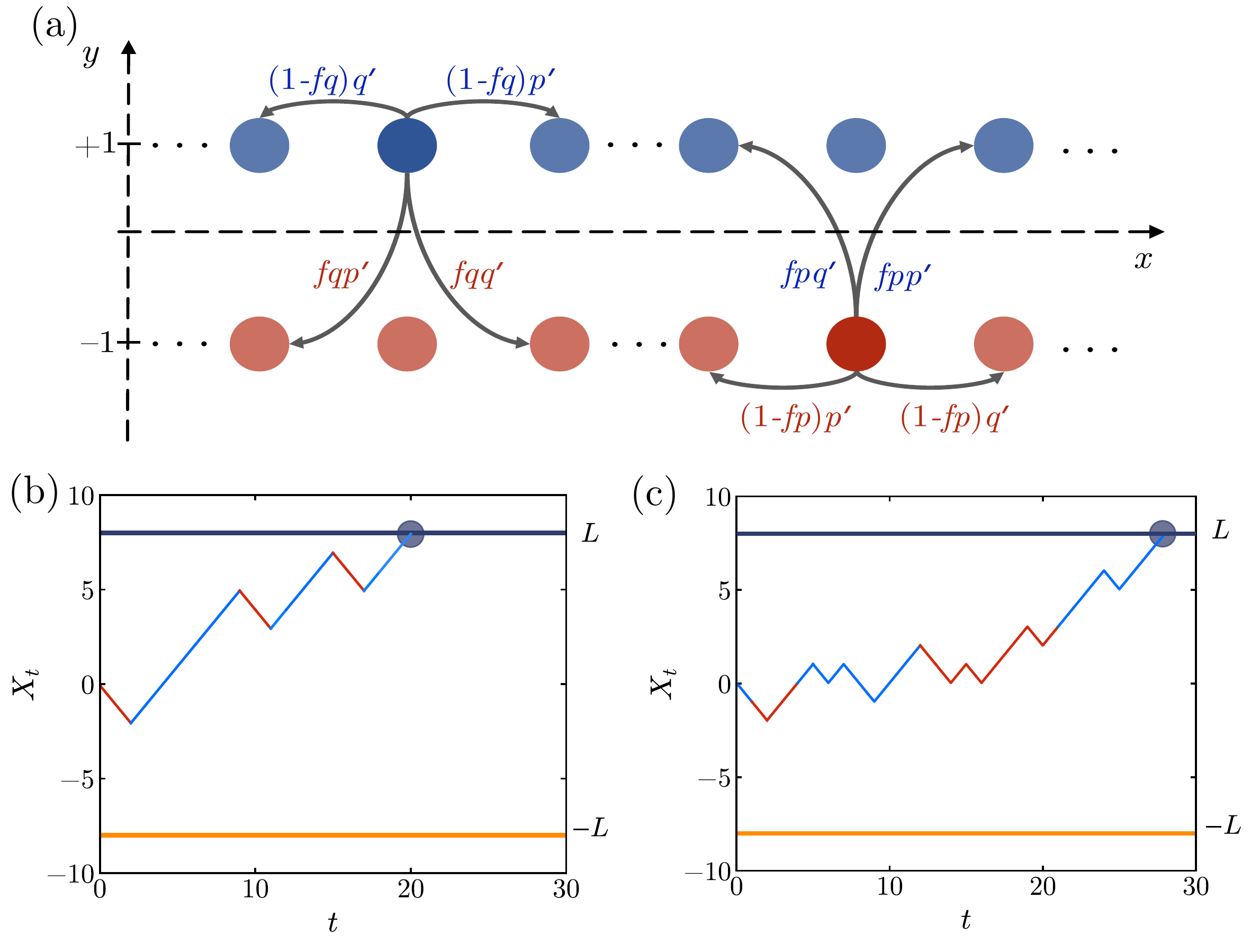}
\caption{(a) Sketch of the RnT model defined by the transition probabilities (\ref{eq:pbeg})--(\ref{eq:pend}). A random walker jumps  at discrete times between states represented as lattice sites (red and blue circles), here with open boundary conditions along $x$. Each state is characterized by two variables: its position along the $x$-axis and the internal preferred-direction variable which is indicated by the position along the $y$-axis: $y=+1$ (or loosely `$y=+$') in the upper lane (blue) and $y=-1$ (or loosely `$y=-$') in the lower lane (red).  Here, $f$ is the tumble probability while $p$ and $q=1-p$ are the probabilities that after a tumble $y=+1$ and $y=-1$ respectively. Furthermore, $p'$ and $q'=1-p'$ denote the probabilities of jumps in and against the preferred direction, with the preferred direction as right in the upper lane and left in the lower lane. (b,c) Sample RnT trajectories with segments shown in different colours to indicate preferred direction: blue for $Y_t=+1$ and red for $Y_t=-1$. In (b) effective tumbles are visible ($p'=1,p=0.6,f=0.25$); movements to the right (left) occur only when the particle is in the upper (lower) lane, which makes the preferred direction identifiable from the particle motion. In (c) effective tumbles are hidden ($p'=0.75, p=0.7, f=0.25$); a particle may move to the right or left in any lane, which renders the preferred direction a genuinely hidden variable.}
\label{f:2lane}
\end{figure}

One way to visualize the present RnT model  is to think of it as a two-lane Markov chain as shown in figure~\ref{f:2lane}(a).  
In addition to the position random variable $X_t$, the state at time~$t$ is now characterized by a second random variable $Y_t$ which has a binary distribution, taking values $y \in \{+,-\}$ to indicate the preferred direction.  When $y=+$, the particle is in the top lane and the transition probabilities to jump to neighbouring sites are given by (with obvious notation for the conditional probability mass functions)
\begin{eqnarray}
    P(x+1,+| x,+) &= (1-fq)p',\label{eq:pbeg}\\
    P(x-1,+| x,+) &= (1-fq)q',\\
    P(x+1,-| x,+) &= fqq',\\
    P(x-1,-| x,+) &= fqp'.
\end{eqnarray}
When $y=-$, on the other hand, the particle is in the bottom lane and the transition probabilities are
\begin{eqnarray}
    P(x+1,-| x,-) &= (1-fp)q',\\
    P(x-1,-| x,-) &= (1-fp)p',\\
    P(x+1,+| x,-) &= fpp',\\
    P(x-1,+| x,-) &= fpq'.\label{eq:pend}
\end{eqnarray}
Since $p+q=p'+q'=1$, the above update rules ensure the normalization $\sum_{x'}\sum_{y'}P(x',y'| x,y)=1$ for all $x\in \mathbb{Z}$ and $y \in \{+,-\}$.  As well as the spatial initial condition (typically $X_0=0$), to apply these rules one also needs to give an initial condition on the preferred direction: we can here make the natural assumption that it starts positive with probability $p$ and negative with probability $q$ (equivalent to having a `pre-tumble' at time zero).

Without loss of generality, we may restrict ourselves to the parameter space $p' \geq q'$ and $p \geq q$, noting that for $p' > q'$ and $p > q$ there is a positive bias, i.e., towards the upper boundary.  The representative trajectories and the distributions in figure~\ref{f:traj} were generated from this model with $p=0.7$, $p'=0.75$, and $f=0.05$.  We see from the trajectories in figure~\ref{f:traj}(a) that, in contrast to an ordinary RW model, the inclusion of the preferred-direction variable leads to temporal correlations: there are time intervals where the particle moves predominantly in the positive direction and time intervals where the particle moves predominantly in the negative direction.   However, except in the persistent limit where $p'=1$ (or, of course,  $p'=0$), it is impossible to infer with certainty the value of the preferred-direction variable at any given time. The distinction between $p'=1$ and $p'\neq 1$ is illustrated in figure~\ref{f:2lane}(b) and (c); the hidden property in the latter case will be important later (see section~\ref{s:results}). As we have already noted, the conditional distributions for $p'\neq1$ in figure~\ref{f:traj}(b) are similar but \emph{not} identical; our aim is ultimately to quantify and understand the statistical difference between them. 

\subsection{Large deviations}

The quantity $X_t/t$ can be thought of as a time-averaged current (with $X_t$ the time-integrated current in a time interval of duration $t$). In the absence of absorbing boundaries, its distribution generically has a large deviation form, loosely stated as
\begin{equation}
\mathrm{Prob}\left (X_t / t =j\right) \sim e^{-I(j)t}, \label{e:ldp}
\end{equation}
with $\sim$ indicating logarithmic asymptotic equivalence (here in the large-$t$ limit) and $I(j)$ being the so-called `rate function'.  The rate function is linked by Legendre transform to the `scaled cumulant generating function' (SCGF):
\begin{equation}
\mu(s)=\lim_{t \to \infty}\frac{1}{t}\log \langle e^{-s X_t} \rangle = -\min_j \left\{I(j)+s j\right\}, \label{e:mu}
\end{equation}
where `$\log$' here, and throughout the text, denotes the natural logarithm. Note that $G_t(s)=\langle e^{-s X_t} \rangle$ is the `moment generating function' for the current up to time $t$; such generating functions play a key role in what follows.

With our sign convention, the Legendre transform maps the region (or `section') of $I(j)$ where the argument $j$ is positive to the region of $\mu(s)$ which has \emph{negative} slope.  Similarly, negative values of $j$ correspond to the section of $\mu(s)$ which has \emph{positive} slope.
The celebrated Gallavotti-Cohen fluctuation relation~\cite{gallavotti1995dynamical} connects the probabilities of positive and negative current fluctuations under quite general conditions and is reflected in a particular symmetry at the level of large deviations: 
\begin{equation} \label{e:GC}
    I(-j)=I(j)+cj, \qquad \mu(s)=\mu(c-s),
\end{equation} with $c$ a constant.   We will refer to this simply as \emph{GC symmetry}.

For the simple RW model, one readily finds that the rate function is given by
\begin{equation}
I_\mathrm{RW}(j) = \frac{1+j}{2}\log \left(\frac{1+j}{2p}\right) + \frac{1-j}{2}\log \left(\frac{1-j}{2q}\right).
\end{equation}
This follows as a standard, albeit tedious, exercise either by starting from the binomial distribution of right (left) steps and using Stirling's formula for the asymptotics, or by noting that since the steps are independent and each has moment generating function $G_\mathrm{RW}(s)=pe^{-s}+qe^{s}$ one has directly $G_t(s)=[G_{RW}(s)]^t = (pe^{-s}+qe^{s})^t$ and so
\begin{equation}
\mu_\mathrm{RW}(s)=\log(pe^{-s}+qe^{s}). \label{e:SCGF_RW}
\end{equation}

For the RnT model, in contrast, it is hard to obtain an explicit analytical expression for the rate function but the SCGF is amenable to calculation either by extending the state space to include the preferred-direction variable and calculating $\mu_\mathrm{RnT}(s)$ as the principal eigenvalue of a tilted generator (see, e.g.,~\cite{TOUCHETTE20091} for more on this general approach) or by exploiting the renewal property of the dynamics, see section~\ref{ss:renewal} and appendix~\ref{s:PSarg}.  In either case one gets
\begin{eqnarray}
\fl \mu_\mathrm{RnT}(s) = \log\frac{1}{2}\bigg\{ & (1-f_+)G_-(s)+(1-f_-)G_+(s)  \nonumber \\
  &+\sqrt{[(1-f_+)G_-(s)+(1-f_-)G_+(s)]^2-4(1-f)G_+(s)G_-(s)}\bigg\}, \label{e:SCGF_RnT}
\end{eqnarray}
where we have defined $f_-=fq$, $f_+=fp$ (the probabilities of effective tumbles conditioned on starting with the preferred direction respectively positive and negative) and the functions $G_\pm(s)=p'e^{\mp s}+q'e^{\pm s}$ (identical to $G_\mathrm{RW}(\pm s)$ with $p$ and $q$ replaced by $p'$ and $q'$).

The SCGFs~\eqref{e:SCGF_RW} and~\eqref{e:SCGF_RnT} provide full information on the asymptotic (large-time) fluctuations in a fixed-time ensemble.  Significantly for the present discussion, it can be shown that, for generic Markov processes,  a large deviation principle in this fixed-time ensemble implies a large deviation principle in a fixed-displacement (i.e., first-passage) ensemble~\cite{gingrich2017fundamental}. In other words, the distribution of FPTs to a \emph{single} barrier a distance $L$ from the starting point has (for large $L$) the form
\begin{equation}
  \mathrm{Prob}\left (T / L =\tau \right) \sim e^{-\phi_\pm(\tau)L}
\end{equation}
where the rate function is $\phi_+(\tau)$ when the barrier is at $+L$ and $\phi_-(\tau)$ when the barrier is at $-L$.  Moreover, these rate functions are related to $I(j)$ of the fixed-time ensemble by
\begin{equation}
    \phi_\pm(\tau) = \tau I \left(\pm 1/\tau \right). \label{e:equiv}
\end{equation}
Equivalently, the SCGFs $\lambda_\pm(s)$ in the fixed-displacement ensemble can also be obtained as
\begin{equation}
\lambda_\pm(s)=\mu^{-1}_\pm(s), \label{e:equiv2}
\end{equation}
where $\mu_+(s)$ is the section of $\mu(s)$ with negative slope, while $\mu_-(s)$ is the section with positive slope. More details on this correspondence can be found in~\cite{gingrich2017fundamental}; note that although that work focuses on Markov jump processes, the proofs seem only to rely on a renewal property of the dynamics. 

Although we are chiefly interested in the case of two absorbing barriers (at $\pm L$) rather than one, we will see shortly that these large deviation relationships still provide valuable information about the presence or absence of the FPT duality in the asymptotic (large-$L$) limit. Similar arguments for wide-interval cases are made in the recent work~\cite{Neri_2025}.

\subsection{Quantifying first-passage-time (FPT) asymmetry}
\label{s:asymmetry}

For the analysis that follows we crucially need a way to measure the degree to which symmetry is broken; we here discuss some options.

\subsubsection{Kullback-Leibler divergence}
One obvious candidate to quantify the asymmetry in FPT statistics (i.e., the breaking of duality) is the Kullback-Leibler Divergence~(KLD)~\cite{Kullback_1951} which is the relative entropy between the conditional distributions for absorption at symmetrically-placed upper and lower boundaries.  For the present discrete case we consider the KLD
\begin{equation}
D_\mathrm{KL} (P_- || P_+) = \sum_{\mathcal{T}>0} P_-(\mathcal{T}) \log \left( \frac{P_-(\mathcal{T})}{P_+(\mathcal{T})} \right).
\label{eq:kld}
\end{equation}
We note that the KLD is not symmetric under interchange of the two distributions but is always zero when they are identical. The KLD is a powerful tool in the present context: zero KLD implies that the FPT duality holds; conversely non-zero KLD signals its violation. 

Evaluation of the KLD is numerically highly challenging due to statistical biases when distributions are poorly sampled; in particular, there are issues with `zero bins' (i.e., insufficient statistics) in the histograms, and this can make it an unreliable measure in practice~\cite{roldan2012entropy}.  However, many estimation techniques have been proposed within the last decade in the non-equilibrium physics literature to provide ways of lower bounding the KLD (see, e.g., chapter~3 in \cite{roldan2014irreversibility}); we discuss some related ideas below.

\subsubsection{Bounds and approximations}
For wide intervals we can use the large deviation forms of the distributions and employ Kullback's inequality~\cite{Kullback_1954} to obtain a bound on the KLD:
\begin{equation}
D_\mathrm{KL} (P_- || P_+) \geq L \phi_+(\bar{\tau}_-), \label{e:Kullback}
\end{equation}
where $\phi_{+}$ is the rate function for absorption at the upper boundary while $\bar{\tau}_-$ is the scaled mean time to absorption at the lower boundary.   We note that in the presence of asymmetry such that $\bar{\tau}_- \neq \bar{\tau}_+ $ then $\phi_+(\bar{\tau}_-)$ is generically non-zero and the KLD is expected to grow as $L$ for large $L$.

A Gaussian approximation for the right-hand side of \eqref{e:Kullback} about the mean yields
\begin{equation}
 L \phi_+(\bar{\tau}_-)  \approx L \frac{(\bar{\tau}_+ - \bar{\tau}_- )^2}{2\sigma^2_+}.\label{e:KLDapprox}
\end{equation}
Although FPT distributions are typically strongly non-Gaussian (especially in the tails), we argue that this approximation is useful in general as an indication of asymmetry. In fact, the right-hand side of~\eqref{e:KLDapprox} is closely related to the (square of) the linear correlation coefficient between the random variables $B$ and $T$; clearly if the first-passage duality holds then measuring the absorption time gives no information about which boundary the particle has been absorbed at and the correlation is zero. In terms of the original unscaled variables,~\eqref{e:Kullback} and~\eqref{e:KLDapprox} motivate us to use the following signal-to-noise ratio
\begin{equation}
    {\rm SNR} = \frac{(\langle T \rangle_+ - \langle T \rangle_- )^2}{2\mathrm{Var}_+(T)} \label{e:hatrho}
\end{equation}
as a simple robust measure of asymmetry in much of our numerical work. In passing, we remark that similar relations between KLD, inequalities, and bounds, have also been discussed by Dechant and Sasa in a more general non-equilibrium context~\cite{Dechant20}.

The approximation $\rm{SNR}$ is numerically more stable than the KLD, in particular because it requires only the first moment of the distribution in the more unlikely direction.  However, $\rm{SNR}$ does not capture the full asymmetry since it compares only the first moments of the distributions to the two boundaries.  In a similar spirit, we may further quantify the asymmetry by comparing higher-order moments through the square of the difference between the $n$th moments scaled by the $n$th power of the FPT variance to the positive (more likely) boundary:
\begin{equation}
    \hat{\rho}_n = \frac{\left(\langle T^n \rangle_+ - \langle T^n \rangle_-\right)^2}{2\mathrm{Var}_+(T)^n}.
    \label{eq:moments_difference}
\end{equation}
Obviously, for $n=1$ this quantity reduces to ${\rm SNR}$ as previously defined.  Knowledge of $\hat{\rho}_n$ for larger $n$ then provides a more detailed characterization of the FPT asymmetry; higher moments, related to skewness and kurtosis ($n = 3$ and $n = 4$, respectively) provide insights into the shape and tail behaviours of the distributions.

Before a detailed analytical and numerical examination of FPT statistics in the RnT model (using the tools developed above), we first devote the next section to considering more generally the conditions under which FPT duality is expected to hold.

%================================================
%================================================
%================================================
\section{Symmetries and renewal processes}
\label{s:symmetry}

\subsection{Martingale approach}
One way to prove the FPT duality in Markov processes uses the mathematical properties of martingales, see~\cite{roldan2023martingales} for a recent tutorial review.  The core of the argument is to define the stochastic entropy production $S_t$ associated with a given trajectory as the logarithm of the probability of that trajectory divided by the probability of its time reversal, then note that the exponential of the negative entropy production $e^{-S_t}$ is a martingale and apply different versions of Doob's optional stopping theorem for martingales.  We refer readers to section~7.4.3 in~\cite{roldan2023martingales} for a detailed mathematical proof of the duality for the entropy production associated with time-homogeneous Markovian stochastic processes in continuous space.

The martingale approach establishes rigorously an FPT duality for the entropy production $S_t$; for Markov processes, current-like quantities are often proportional to entropy production (at least up to boundary terms) so one also gets a first-passage duality for currents (at least asymptotically).  As previously noted, for single-particle processes the position is an integrated current (given by the net number of jumps in the positive direction) so, in many cases, the FPT duality for position follows directly~\cite{neri2017statistics}. Similar considerations hold for unicyclic continuous-time Markov processes with discrete states, for which the net number of cycles is proportional to the entropy production and thus also satisfies the FPT duality. Such a symmetry property in the context of chemical reactions has been thoroughly analyzed under the name of generalized Haldane equality~\cite{qian2006generalized,ge2012stochastic}.

The qualitative argument that the symmetry is broken when time-integrated currents are not proportional to entropy production is illustrated by the fact the FPT duality does not hold in general for the position of a driven Brownian particle in a non-linear potential or for the net number of cycles in a multi-cyclic Markov jump model. More detailed treatment of the symmetry violations in different contexts is provided in the very recent contributions by Piephoff and Cao~\cite{piephoff2025} and by Neri~\cite{Neri_2025}; in a similar spirit to the latter work we now aim to provide further insights by considering intuitive arguments based on symmetry properties of trajectories and the Galavotti-Cohen fluctuation symmetry. 

\subsection{Trajectory properties for Markovian processes}
\label{ss:traj}

\begin{figure}
\centering
\includegraphics[scale=0.97]{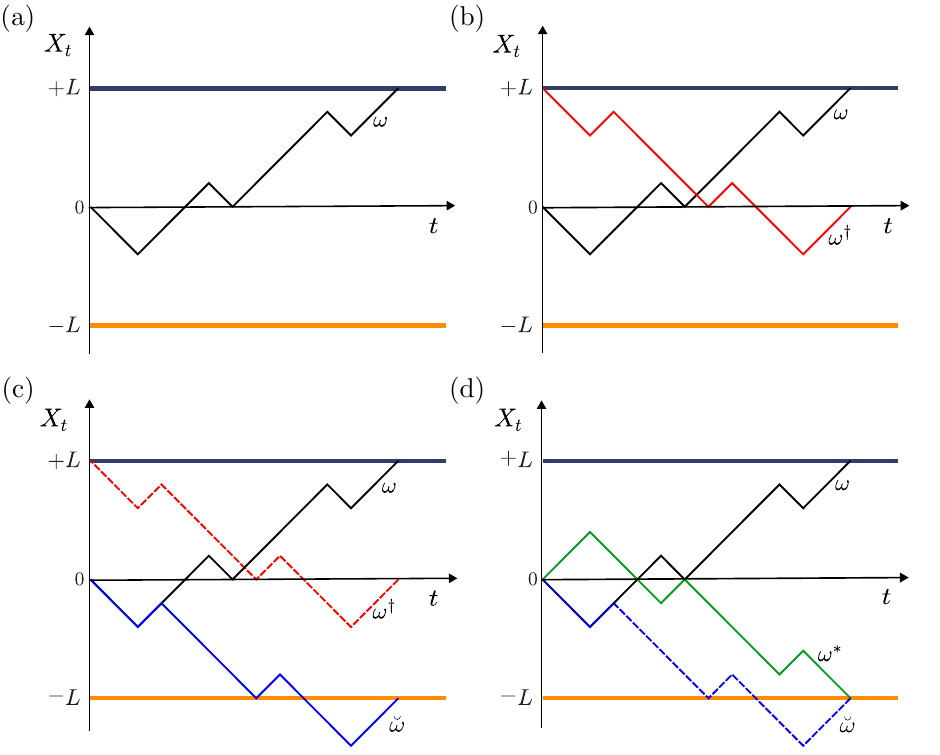}
\caption{Probability-preserving operations on a representative RW trajectory: (a) original trajectory, (b) time reversal, (c) spatial translation, (d) step reordering.}
\label{f:trajs}
\end{figure}

Indeed, for simple random walks (like our RW model of section~\ref{sss:RW}), one can see the duality directly by considering the symmetry properties of trajectories.  The heuristic argument is sketched here with reference to example trajectories in figure~\ref{f:trajs}.

We start by considering a particular RW trajectory $\omega$ starting at the origin and first reaching the upper boundary at time $\T$ as shown in figure~\ref{f:trajs}(a).  By construction, this trajectory has $X_0=0$ and $X_\T=L$.  We denote its probability mass function by $P(\omega)$.  Next we look at the time reversal of this trajectory denoting it by ${\omega}^\dagger$ and noting that we have ${X}^\dagger_0=L$ and ${X}^\dagger_\T=0$, see figure~\ref{f:trajs}(b).  The probability of this trajectory under the same RW dynamics but conditioned on starting at $L$ is denoted by $P^\dagger({\omega}^\dagger)$.  Comparing probabilities one immediately sees
\begin{equation}
\frac{P({\omega})}{P^\dagger({\omega}^\dagger)}= \left( \frac{p}{q} \right)^L = e^{L \log(p/q)}. \label{e:timesym}
\end{equation}
We remark that the quantity $L\log(p/q)$ appearing in the exponent here is the (environment) entropy production associated with the  trajectory $\omega$ and indeed we see that this entropy is proportional to final position; we can write~\eqref{e:timesym} in the form
\begin{equation}
\frac{P(\omega)}{P^\dagger(\omega^\dagger)}=e^{cL},\label{e:timesym2}
\end{equation}
where the constant $c$ is the entropy production for a single forward step.  Such a time-reversal relation is closely related to the GC symmetry~\eqref{e:GC}, a point to which we will return later.\footnote{More generally, one can endow both forward and reverse processes with an initial distribution over sites, leading to the possibility of an $L$-independent boundary term in the exponent on the right-hand side of~\eqref{e:timesym2}; if the initial distributions are identical then of course we can write simply $P(\omega^\dagger)$ in the denominator on the left-hand side.}

To see the connection to the FPT duality, we start by noting that the trajectory ${\omega}^\dagger$ (conditioned on starting at $L$) has identical probability to the spatial translation $\breve{\omega}$ shown in figure~\ref{f:trajs}(c), which is conditioned to start at $\breve{X}_0=0$ and ends at $\breve{X}_\T=-L$; space-homogeneity enforces
\begin{equation}
    P(\breve{\omega})=P^\dagger(\omega^\dagger).\label{eq:eg0}
\end{equation}
Note that the trajectory $\breve{\omega}$ is not necessarily a first-passage trajectory to $-L$; as shown in figure~\ref{f:trajs}(c), it may have first crossed $-L$ at an earlier time than $\T$.  However, since the random walk has identical independent steps we can rearrange their order in $\breve{\omega}$ to create a trajectory $\omega^*$ which reaches $-L$ for the first time at $\T$  and has the same probability; time-homogeneity enforces  
\begin{equation}
    P(\omega^\star)=P(\breve{\omega}).\label{eq:eg1}
\end{equation}
In particular, we can always do this in such a way that the new trajectory $\omega^*$ is just a spatial reflection of the original trajectory $\omega$, see~\fref{f:trajs}(d).  In other words, for every first-passage trajectory $\omega$ with $X_\T=L$ we have a one-to-one mapping to a first-passage trajectory $\omega^*$ with $X_\T=-L$; the probabilities of these two trajectories are related by
\begin{equation}
\frac{P(\omega)}{P(\omega^*)}= \left( \frac{p}{q} \right)^L = e^{L \log(p/q)},~\label{e:ratio}
\end{equation}
which follows from~\eqref{e:timesym}, \eqref{eq:eg0}, and~\eqref{eq:eg1}. 
The crucial observation is that the probability ratio in~\eqref{e:ratio} depends only on the jump probabilities and the barrier position: it is independent of the value of $\T$.  This is essentially just the requirement needed in the classical method of images where for every first-passage trajectory to an absorbing boundary we must have a one-to-one mapping to a reflected trajectory (starting at the image) with the ratio of these trajectory probabilities independent of the time of absorption, see e.g.,~\cite{redner}. Summing over all trajectories with $T=\T$ we thus get
\begin{equation}
\mathrm{Prob}(T=\T,B=+1)=e^{L \log(p/q)}\mathrm{Prob}(T=\T,B=-1),~\label{e:joint}
\end{equation}
using again the scaled  variable $B=X_T/L$.  Further summing both sides of~\eqref{e:joint} over $\T$ we find the `fluctuation relation' between the splitting probabilities
\begin{equation}
\frac{\mathrm{Prob}(B=+1)}{\mathrm{Prob}(B=-1)}=e^{L \log(p/q)}, \label{e:split2}
\end{equation}
as also emerges straightforwardly from the martingale approach.  Finally, combining~\eqref{e:joint} with~\eqref{e:split2} yields equality of the conditional probability distributions, i.e., the fundamental FPT duality $P_+(\mathcal{T})=P_-(\mathcal{T})$ given by~\eqref{e:duality}.

At this point, we digress to note that the same trajectory-symmetry arguments also hold between {\em two}  related single-barrier problems, one with an absorbing threshold at $L$ and another with the absorbing threshold at $-L$.  We let $P^{(1)}_+(\mathcal{T})$ be the probability that our RW model starting at the origin is absorbed at time $\T$ at a barrier at $+L$ in the case where there is no barrier (absorbing or otherwise) at $-L$.  Similarly, we let $P^{(1)}_-(\mathcal{T})$ be the probability that our RW model starting at the origin is absorbed at time $\T$ at a barrier at $-L$ when there is no barrier (absorbing or otherwise) at $+L$.  Note that these two probabilities now correspond to two different physical set-ups in which the barrier position is reflected through the origin (or, equivalently, the barrier position is fixed but the field direction is reversed).  There are obviously more trajectories that contribute to $P^{(1)}_+(\mathcal{T})$ than to $P_+(\mathcal{T})$.  However, repeating the operations of figure~\ref{f:trajs} one still finds the one-to-one mapping between trajectories contributing to $P^{(1)}_+(\mathcal{T})$ and those contributing to $P^{(1)}_-(\mathcal{T})$ and hence the duality also holds here: $P^{(1)}_+(\mathcal{T})=P^{(1)}_-(\mathcal{T})$.

As shown above, the duality for the simple RW model is rather intuitive; indeed one might ask why the argument didn't simply start from the reflected trajectory of figure~\ref{f:trajs}(d).  We contend here that the trajectory transformations in the other subfigures reveal equivalences which give useful insight into the underlying structure of the symmetry and reveal the connection to time reversal and the GG symmetry.  This is useful in more complicated cases where it may be less obvious how to calculate the relative probabilities of a trajectory and its reflection.  In particular, we note that the trajectory-symmetry argument does not rely on a full global rearrangement to the reflected trajectory in figure~\ref{f:trajs}(d); in fact, we need only a minimal local rearrangement to convert a non-first-passage trajectory into a first-passage trajectory.

It is straightforward to see that any single-particle random walk model, with independent identically-distributed (i.i.d.) step lengths, which obeys a time-reversal symmetry of the form~\eqref{e:timesym2} will automatically satisfy the FPT duality.  Furthermore, to check the time-reversal symmetry, we need only look at \emph{one-step} probabilities encoded in the moment generating function for a single step $G(s)$: since $G_t(s)=[G(s)]^t$ in the i.i.d.\ case, if we have $G(s)=G(c-s)$ for some $c$, then~\eqref{e:timesym2} will hold and hence the FPT duality for position. Of course, it is easy to construct Markov processes where a relation of the form~\eqref{e:timesym2} does not hold. For example, for random walk models with particular asymmetric distributions of step lengths one does not have $G(s)=G(c-s)$ and there is thus no FPT duality in either the original two-boundary problem or the variant single-boundary problem.  In general, the nature of this duality violation will be different in the two problems (quantified, for example, using the measures in section~\ref{s:asymmetry}, where for the single-boundary case the $+$ and $-$ subscripts refer to the bias being towards or away from the boundary respectively).  However, the relative number of trajectories contributing to $P^{(1)}_+(\tau L)$ but not $P_+(\tau L)$ is expected to decrease as $L$ increases with fixed $\tau$. Loosely speaking, for large $L$ we therefore anticipate $P^{(1)}_+(\T) \approx {P}_+(\T)$ and $P^{(1)}_-(\T) \approx {P}_-(\T)$, as suggested e.g., by~\cite{saito2016waiting}.

Going beyond i.i.d.\ steps, one finds in some models that~\eqref{e:timesym2} does not hold exactly for finite $L$ but equality is restored in the large-$L$ limit corresponding to the fluctuation relation in its usual asymptotic form (i.e., the GC symmetry); this is the situation, for example, for a simple random walk with hopping probabilities which are periodic in space. In such cases, transformations like those in figures~\ref{f:trajs}(c) and~\ref{f:trajs}(d) generically also yield `finite-size' corrections to the trajectory probabilities;  significantly, we note that for any process with short-range correlations in time we should be able to carry out the minimal rearrangement required to obtain a first-passage trajectory in figure~\ref{f:trajs}(d) with only a sub-leading correction to the probability.  Hence one finally gets $P(\omega)/P(\omega^*)=e^{cL+c_b}$, where the boundary term $c_b$ is sub-extensive in $L$, and so the duality is restored asymptotically for large $L$.  

Pulling this subsection together, we argue heuristically that if the GC symmetry relation holds (in the large-$t$ limit) for an integrated current associated with a given Markov process, then we expect the first-passage duality in the large-$L$ limit for \emph{both} the two-boundary problem and the single-boundary variant.  Indeed the claim for the single-boundary case was already rigorously shown by Gingrich and Horowitz~\cite{gingrich2017fundamental}: it follows directly from~\eqref{e:equiv} that the GC symmetry $I(-j)=I(j)+cj$ in the fixed-time ensemble implies $\phi_-(\tau)=\phi_+(\tau)+c$ in the fixed-displacement ensemble which is a statement of the  duality in the asymptotic large-$L$ limit.

\subsection{Non-Markovian processes and renewals}
\label{ss:renewal}

In the previous subsection, we gave trajectory-based arguments for (asymptotic) FPT duality in time-homogeneous Markov processes. For non-time-homogenous or non-Markovian processes, the situation is more complicated; in particular, for systems with memory one cannot generically expect to rearrange current increments (steps in position space) with only a sub-leading correction to the trajectory probability.   The connection of trajectory reversal to entropy production may also be less immediately clear, as is the case for active processes, such as the RnT model.  For the RnT with $0<p'<1$ then observing the spatial trajectory is not enough to infer the preferred-direction (spin) variable.  If the latter is known then we can define two different entropies: one in which the preferred direction is regarded as an even-parity variable (corresponding to considering a Markov process with the enlarged state space of position and preferred direction) and one in which the preferred direction is regarded as an odd-parity variable, with signs ﬂipped in the reversed trajectory.  In general, neither of these entropies is proportional to the position so martingale properties of the entropy production do not imply the first passage duality. For further discussion on entropy and irreversibility in active matter, see, e.g.,~\cite{ShankarMarchetti,PhysRevX.9.021009,pruessner2025fieldtheoriesactiveparticle}.

In this subsection we focus on the special case of processes with a renewal structure, as accessible examples of models which are non-Markovian in position space. We present here the mathematical framework and physical intuition which allow determination of whether FPT duality holds in models of this class before testing the resulting predictions in section~\ref{s:results} against analytical and numerical results for the RnT.  

\subsubsection{Finite-time renewal structure}
The general construction is that the durations of times between renewal events (i.e., the lengths of renewal intervals) are i.i.d.\ random variables $N_i$ ($i\geq 1$) from some discrete probability distribution with strictly positive support; we will later want to require that all moments of the distribution are finite.  We denote by $R_t$ the number of renewals up to time step $t$, excluding the (implicit) renewal at time $t=0$. Building on the renewal process $R_t$,  we can represent the integrated current for many models in terms of an associated renewal-reward process.   For example, in a single-particle model (like the RnT of section~\ref{ss:RnT}) where the integrated current is just position we can write
\begin{equation}
    X_t= \left(\sum_{i=1}^{R_t} \Delta^{(i)} \right)+ \Delta^{(F)} ,\label{e:renewalreward}
\end{equation}
where $\Delta^{(i)}$ is the change in position in the $i$th renewal interval (and can be thought of as a terminal `reward' applied at the end of the interval) while $\Delta^{(F)}$ is the position increment from the residual interval between the last renewal and time $t$.  The $\Delta^{(i)}$s may of course be positive or negative (and will typically depend on the lengths of the corresponding renewal intervals); the crucial point for the renewal analysis is only that they are i.i.d.\ conditioned on $N_i$.  The final increment $\Delta^{(F)}$ will generally be drawn from a different distribution.

The sharp-eyed reader may notice that in the above discussion we have been rather vague about the definition of a renewal event.  In fact, for a given model there may be more than one way to decompose the current in the form of~\eqref{e:renewalreward}. In the RnT, for example, we could identify each tumble as a renewal, or every second tumble.  The latter is perhaps unnecessarily complicated but it does make sense to consider every second \emph{effective} tumble (i.e., split the trajectory only at points where the preferred direction changes from positive to negative, or only at points where it changes from negative to positive, see again section~\ref{ss:RnT} for more on the notion of effective tumbles) so that the current in each renewal interval is the sum of a portion with positive preferred direction and a portion with negative preferred direction. On the other hand, we cannot identify every effective tumble as a renewal since then the current increments would not have the required i.i.d.\ property.

A central object in the symmetry analysis is the weighted moment generating function $W_n(s)$ for a single renewal interval of length $n$; to be concrete this is the usual moment generating function for integrated current in a renewal interval of $n$ time steps multiplied by the probability of seeing a renewal interval of that length, i.e., $W_n(s)=\mathrm{Prob}(N_i=n)\langle e^{-s\Delta^{(i)}} \rangle_n$ where the subscript $n$ indicates that the expectation depends on $n$ (see appendix~\ref{s:PSarg} for some specific examples). Now, if we have the generating-function symmetry $W_n(s)=W_n(c-s)$ for a single renewal interval, then we will obviously have it for any number of them.  This gives the GC symmetry for any complete number of renewal intervals which is related to the idea of looking at fluctuation relations at `beat' times~\cite{Garilli_2024}.  Equivalently one can ask whether the integrated current (position) is proportional to an entropy-like quantity after a complete number of renewal intervals; if so, we have martingale properties \emph{at those beat times}. However, first-passage will not typically occur after an exact number of renewal intervals so in general we do not expect the FPT duality to hold for finite system sizes.  Of course, for a given model, one can (in principle) explicitly check the statistics from the final incomplete renewal interval; rather than performing such a calculation, we now turn to the large-system limit, anticipating that details of the final incomplete interval are not relevant for determining whether FPT duality holds in this case.

\subsubsection{Conditions for asymptotic FPT duality}

Intuitively, we can build up general understanding by again thinking about a trajectory picture.  For the asymptotic behaviour, we only need consider a complete number of renewal intervals, assuming the $N_i$ have finite moments (which excludes heavy-tailed distributions).  In other words, we merely look to see if the barrier has been reached at the renewal times themselves, arguing that any corrections due to hitting the barrier between these times are finite (and only weakly dependent on the barrier position) and so when divided by $L$, they scale to zero in the large-$L$ limit.  Furthermore, due to the renewal property, we can freely change the order of the renewal intervals without changing the statistics of the FPTs, just as we did for the simple RW model in section~\ref{ss:traj}.

From a formal perspective, we again note that the Gingrich-Horowitz proof~\cite{gingrich2017fundamental} for equivalence between GC symmetry (limit of long time) and first-passage duality (limit of large barrier position) apparently requires only a renewal property of the current.  It obviously then follows that for processes with the renewal structure we can just test the GC symmetry to determine whether the first-passage duality holds: if, and only if,~\eqref{e:GC} is obeyed for the current large deviations we have the asymptotic FPT duality.  Obviously, this test is only really useful if one can obtain the large deviation properties of the current which may seem a formidable task.  However, happily, the renewal structure facilitates calculation of the SCGF via a well-established procedure in Laplace space which starts from $W_n(s)$ for a single renewal interval (and assumes that corrections from the last incomplete interval are subleading); see appendix~\ref{s:PSarg} for an outline of the method and details of its application to the RnT model with two different decompositions into renewal intervals.  The result for the RnT is the expression already given in~\eqref{e:SCGF_RnT} and, with $f\in (0,1)$, we see that the GC symmetry~\eqref{e:GC} only holds when either $p$ or $p'$ takes one of the special values 0, 1/2, or 1.\footnote{To be precise, one has the symmetry in the following cases: (i) $p=1/2$ and any $p' \in [0,1]$, (ii) $p=0$ or $p=1$ and any $p' \in (0,1)$, (iii) any $p \in [0,1]$ and $p'=1/2$, (iv) any $p \in (0,1)$ and $p'=0$ or $p'=1$.  Of course, in this paper, we mainly restrict consideration to $p \in [1/2,1]$ and $p' \in [1/2,1]$.}

A natural question to ask is whether we can further leverage the renewal structure to determine the presence or absence of the asymptotic duality from the properties of a single renewal interval, just as we could predict it for a random walk of i.i.d.\ steps from the generating function for a single time step (see section~\ref{ss:traj}).  The answer is affirmative.    After all, if the (weighted) moment generating function for a single renewal interval obeys $W_n(s)=W_n(c-s)$, we have the symmetry for any finite number $m$ of renewal intervals and can take the $m \to \infty$ limit.  Hence, since asymptotically we need only consider a complete number of renewal intervals (again for $N_i$ with finite moments), we trivially obtain that $W_n(s)=W_n(c-s)$ is a \emph{sufficient} condition for the asymptotic duality.  However, it turns out that $W_n(s)=W_n(c-s)$ is not a \emph{necessary} condition for the asymptotic duality.\footnote{Mathematically, the necessary condition is only that $z^*(s)=z^*(c-s)$ where $z^*(s)$ is the value of $z$ which solves $\sum_{n=1}^\infty z^n W_n(s)=1$; see appendix~\ref{ss:PSarg0}.}  For example, consider the RnT model in the special case $p'=1$. If we identify each tumble as a renewal then the weighted moment generating function for a single renewal interval is $W^A_n(s)=f(1-f)^{n-1}[pe^{-sn}+qe^{sn}]$ (see appendix~\ref{ss:PSarg1}) which \emph{does not} satisfy  $W^A_n(s)=W^A_n(c-s)$ although the SCGF characterizing the asymptotic behaviour after many tumbles does obey the GC symmetry~\eqref{e:GC}: in this case, we get $\mu_\mathrm{RnT}(s)=\mu_\mathrm{RnT}(c-s)$ with $e^c=(1-f_-)/(1-f_+)=(1-fq)/(1-fp)$. Physically, this situation seems to be due to the fact that the dynamics is ballistic (chosen from one of two options) in each renewal interval so we need the many-tumble limit to access the full distribution of current.  On the other hand, if we choose renewals at effective tumbles from negative to positive (see appendix~\ref{ss:PSarg2}) then, again with $p'=1$, the generating function for a single renewal interval \emph{does} obey the symmetry as easily seen from its Laplace-transformed form~\eqref{e:tildeW2}.  Here the dynamics in each complete renewal interval is the same (a portion with preferred direction positive, followed by a portion with preferred direction negative).  It seems intuitively plausible that for renewal processes in which the dynamics in each complete renewal interval is identical (meaning here with the same sequence of hidden-variable values), the symmetry $W_n(s)=W_n(c-s)$ is necessary as well as sufficient for asymptotic duality; we have checked this is consistent with models in a restricted class but do not provide a general rigorous proof.

We now summarize the central claims which we seek to demonstrate in the remainder of the paper.  For renewal processes where the inter-renewal times have finite moments then:
\begin{itemize}
    \item If the single-renewal moment generating function symmetry $W_n(s)=W_n(c-s)$ holds, the asymptotic FPT duality certainly holds;
    \item If the single-renewal moment generating function symmetry $W_n(s)=W_n(c-s)$ fails, and dynamics in each renewal interval is identical, than the asymptotic FPT duality is expected to fail. 
\end{itemize}
In particular, in section~\ref{s:results} we will examine the asymptotic behaviour (and finite size corrections) of the asymmetry measures in section~\ref{s:asymmetry} for two specific RnT examples corresponding to the two classes above and seek to further elucidate their physical distinction.  Towards that end, we first present details of the FPT statistics in the RnT model, with the non-trivial analytics meriting a separate section.

%================================================
%================================================
%================================================

%================================================
%================================================
%================================================
%================================================
\section{Analytical FPT results for the RnT model}
\label{s:fpt-theory}

In this section, we derive exact analytical expressions for the first-passage statistics of the RnT model, as defined in section~\ref{ss:RnT}, with two absorbing boundaries. 

A series of exact analytical results have recently emerged regarding first-passage statistics for discrete-time dynamics of both Markovian~\cite{giuggioli2020,sarvaharmangiuggioli2020,dasgiuggioli2022,giuggiolisarvaharman2022,dasgiuggioli2023,marris2023exact,das2023misconceptions,barbini2024lattice,giuggioli2024multi} and non-Markovian~\cite{larralde2020first,marris2024persistent} lattice random walks. Specifically, the first-passage statistics can be obtained by exploiting the classical Montroll's defect technique~\cite{montrollwest1979,montroll1956,montrollweiss1965,kenkre2021a} which uses the linearity of the master equation and the knowledge of the defect-free propagator (i.e., the propagator in the absence of absorbing sites). 
Here, we undertake a technically demanding adaptation of the defect technique to the asymmetric run-and-tumble process with two absorbing boundaries.
In this approach, we first solve for the defect-free propagator of our RnT model and then use it to obtain the FPT probabilities in the presence of two absorbing boundaries.
To provide general results for future work, and align our calculations with the literature mentioned above, we first treat here arbitrary initial conditions before specializing to those required to investigate the FPT duality.

\subsection{Propagator transforms}

We start by considering the probability $\mathrm{Prob}(X_t=x,Y_t=\pm)$ of finding the run-and-tumble particle (see figure~\ref{f:2lane}(a)) on site $x$ at time $t$ with its `hidden' state set to $\pm$.  We denote this quantity by $P_t(x,\pm)$ but stress that, in contrast to the earlier set-up in section~\ref{s:setup}, we here assume a general initial condition $P_0(x,\pm)$; we will later use the notation of conditioning to indicate probabilities at time $t$ starting from a specific initial site/state.
The discrete-time master equation may be written as
\begin{eqnarray}
    \left( \begin{array}{c}
        P_{t+1}(x,+) \\ P_{t+1}(x,-)
    \end{array} \right)
    &=
    \left( \begin{array}{cc}
        (1-f q)q' & fpq'\\ fqp' & (1-fp)p'
    \end{array} \right)    
    \left( \begin{array}{c}
        P_{t}(x+1,+) \\ P_{t}(x+1,-)
    \end{array} \right) \nonumber \\
    &\quad+
    \left( \begin{array}{cc}
        (1-f q)p' & fpp'\\ fq q' & (1-fp)q' 
    \end{array} \right) 
    \left( \begin{array}{c}
        P_{t}(x-1,+) \\ P_{t}(x-1,-) 
    \end{array} \right) .\label{eq:c1}
\end{eqnarray}
Now, using the discrete Fourier transform given by 
\begin{equation}
    \widehat{P}_t(k,\pm) = \sum_{x=-\infty}^{\infty} e^{i k x} P_t(x,\pm),
\end{equation} one obtains from~\eqref{eq:c1} that
\begin{equation}
    \left( \begin{array}{c}
        \widehat{P}_{t+1}(k,+) \\ \widehat{P}_{t+1}(k,-)
    \end{array} \right)
    = \mathbb{M} 
    \left( \begin{array}{c}
        \widehat{P}_{t}(k,+) \\ \widehat{P}_{t}(k,-)   \label{eq:dft}
    \end{array} \right),
\end{equation}
where
\begin{equation}
\fl    \mathbb{M} =
    \left( \begin{array}{cc}
        (1-f q)q'e^{-i k}+ (1-fq) p' e^{i k} & fpq'e^{-i k}+ fpp' e^{i k}\\ 
        fqp'e^{-i k}+ fqq' e^{i k} & (1-fp)p'e^{-i k}+ (1-fp) q' e^{i k}
    \end{array} \right)    .
\end{equation}
Then, introducing the $z$-transform of $\widehat{P}_t(k,\pm)$ as 
\begin{equation}
    \widetilde{\widehat{P}}_z(k,\pm) = \sum_{t=0}^\infty z^t \widehat{P}_t(k,\pm),
\end{equation}
we can recast~\eqref{eq:dft} in the $z$-domain as
\begin{equation}
  \left( \begin{array}{c}
        \widetilde{\widehat{P}}_{z}(k,+) \\ \widetilde{\widehat{P}}_{z}(k,-)
    \end{array} \right)
    = \big( {\mathbb{I}} - z \mathbb{M} \big)^{ - 1} 
    \left( \begin{array}{c}
        \widehat{P}_{0}(k,+) \\ \widehat{P}_{0}(k,-)
    \end{array} \right)   ,  \label{eq:inv_sol}
\end{equation}
where $\mathbb{I}$ is the $2\times2$ identity matrix.

When the particle starts at $t=0$ from site $x_0$ with state $+$, one has $P_0(x,+) =  \delta_{x,x_0}$, $\widehat{P}_{0}(k,+) =  e^{i k x_0}$, $P_0(x,-) = 0$, and $\widehat{P}_{0}(k,-) = 0$. Consequently, using now the conditional probability notation to explicitly indicate the initial condition, \eqref{eq:inv_sol} yields
\begin{eqnarray}
    \widetilde{\widehat{P}}_{z}(k,+|x_0,+) &= \frac{ e^{i k(x_0+1)} }{D_z(s)}  \left[-e^{2 i k} q' z (1-f p) - p' z (1-f p) + e^{i k} \right] , \label{eq:P+_x0+}\\
    \widetilde{\widehat{P}}_{z}(k,-|x_0,+) &=   \frac{ e^{i k(x_0+1)} }{D_z(s)}  \left[ f q z \left(p'+e^{2 i k} q'\right)\right] , \label{eq:P-_x0+} 
\end{eqnarray}
with 
\begin{eqnarray}
\fl D_z(s) &=  ( e^{4 i k}+1) (1-f)  (1-q') q' z^2 -e^{3 i k} z [1-f q -f (1-2 q) q'] \nonumber \\
\fl & + e^{2 i k} \left[ 1+(1-f) (1-2q'+2 q'^2) z^2 \right]  - e^{i k} [1 - f + f q' + f q (1 - 2 q') ] z, \label{eq:def_Dk}
\end{eqnarray}
where we have used $p+q=p'+q'=1$. Similarly, when the particle starts from $x_0$ with state $-$, one has $P_0(x,+) =  0$, $\widehat{P}_{0}(k,+) =  0$, $P_0(x,-) = \delta_{x,x_0}$, and $\widehat{P}_{0}(k,-) = e^{i k x_0}$. In this case,~\eqref{eq:inv_sol} gives
\begin{eqnarray}
    \widetilde{\widehat{P}}_{z}(k,+|x_0,-) &=  \frac{ e^{i k(x_0+1)} }{D_z(s)} \left[  f p z \left(q'+e^{2 i k} p'\right) \right] , \label{eq:P+_x0-}\\
    \widetilde{\widehat{P}}_{z}(k,-|x_0,-) &=  \frac{ e^{i k(x_0+1)} }{D_z(s)}   \left[-e^{2 i k} p' z (1-f q) - q' z (1-f q) + e^{i k}\right] \label{eq:P-_x0-} .
\end{eqnarray}
The propagator $z$-transforms $\widetilde{P}_z(x,y|x_0,y_0)$ for $y\in\{+,-\}$ and $y_0\in\{+,-\}$ are obtained from the quantities $\widetilde{\widehat{P}}_{z}(k, y | x_0,y_0)$ using the inverse Fourier transform given by
\begin{equation}
    \widetilde{P}_z(x,y|x_0,y_0) = \frac{1}{2 \pi} \int_{-\pi}^{\pi} { e^{- i k x}  \widetilde{\widehat{P}}_{z}(k, y | x_0,y_0) \, \rm{d}}k .
    \label{eq:Prop_in_xz}
\end{equation}

\subsection{FPT statistics}

To solve the FPT problem, we follow the established  procedure of first considering partially absorbing boundaries and then taking the limit of perfect absorption~\cite{giuggiolisarvaharman2022,marris2024persistent}. To this aim, we consider  two partially absorbing sites $L$ and $-L$ such that the particle upon reaching these sites is absorbed with probability $\varrho$ irrespective of its hidden state. For $\varrho \to 1$, these sites reduce to perfectly absorbing boundaries. In the two-lane description of the dynamics (see figure~\ref{f:2lane}(a)), the coordinates of the absorbing sites are described by the site-state pair $(L,y_L)$ and belong to the  set $\mathfrak{S} = \{(L,+),(L,-),(-L,+),(-L,-)\}$ of $|\mathfrak{S}|=4$ elements.  The master equation of the dynamics with the absorbing sites may be written as~\cite{giuggiolisarvaharman2022,marris2024persistent}
\begin{eqnarray}
\fl    P^{(a)}_{t+1}(x,y) = \sum_{x'=-\infty}^{\infty}\sum_{y' = \{+,-\}} &\Big[ P(x,y\vert x',y') P^{(a)}_t(x',y')   \nonumber \\
\fl    &- \varrho \sum_{i=1}^{|\mathfrak{S}|} \delta_{x,L_i} \delta_{y,y_{L_i}}   P(L_i,y_{L_i}\vert x',y') P^{(a)}_t(x',y') \Big] ,
    \label{eq:absorb_master_in_t}
\end{eqnarray}
where $P^{(a)}_t(x,y)$ denotes the probability, in the presence of the absorbing site, of finding the particle on site $x$ at time $t$ with state $y$, while $P(x,y\vert x',y')$ is the one-step transition probability from site-state pair $(x',y')$ to site-state pair $(x,y)$. Note that, here and elsewhere, the sum from $i=1$ to $|\mathfrak{S}|$ denotes a summation over the elements of the set $\mathfrak{S}$ with $(L_i,y_{L_i})$ being the pair of dummy variables such that $(L_1,y_{L_1})=(L,+)$, $(L_2,y_{L_2})=(L,-)$, $(L_3,y_{L_3})=(-L,+)$ and $(L_4,y_{L_4})=(-L,-)$. 

We remark that for $\varrho=0$, i.e., in the absence of absorbing sites,~\eqref{eq:absorb_master_in_t} reduces to the defect-free master equation~\eqref{eq:c1}. By taking $\varrho \sum_{i=1}^{|\mathfrak{S}|} \delta_{x,L_i} \delta_{y,y_{L_i}}   P(L_i,y_{L_i}\vert x',y') P^{(a)}_t(x',y')$ as a known function, the formal solution of~\eqref{eq:absorb_master_in_t} is simply a sum of the defect-free propagator, i.e., the Green’s function solution of the defect-free master equation~\eqref{eq:c1}, and that same propagator convoluted in time and space with the known term. To be concrete, one may write the desired solution as
\begin{eqnarray}
\fl    P^{(a)}_{t}(x,y) &= \sum_{x'}\sum_{y'}  P_t(x,y|x',y') P^{(a)}_0(x',y')  \nonumber \\
\fl    & \quad - \sum_{t'=0}^{t-1} \varrho \sum_{i=1}^{|\mathfrak{S}|}  P_{t-1-t'}(x,y|L_i,y_{L_i}) \sum_{x'}\sum_{y'} P(L_i,y_{L_i}\vert x',y') P^{(a)}_{t'}(x',y') 
    \label{eq:P(a)t_xy},
\end{eqnarray}
where for brevity we have omitted the limits of the sums over $x'$ and $y'$. Here, $P_t(x,y|x',y')$ is the defect-free propagator with $P_0(x,y\vert x',y')=\delta_{x,x'} \delta_{y,y'}$; note that $P_1(x,y|x',y')$ is just the one-step transition probability $P(x,y\vert x',y')$. The second term on the right-hand side of~\eqref{eq:P(a)t_xy} considers the probability leakage due to all possible trajectories that last visited one of the absorbing sites at time $t'+1$, for $t'<t$.
In the $z$-domain, the solution~\eqref{eq:P(a)t_xy} yields
\begin{eqnarray}
\fl    {\widetilde{P}}^{(a)}_{z}(x,y)  =\sum_{x'}\sum_{y'}  \Big[& {\widetilde{P}}_z(x,y|x',y') P^{(a)}_0(x',y')   \nonumber \\
\fl    &- z  \varrho \sum_{i=1}^{|\mathfrak{S}|} {\widetilde{P}}_{z}(x,y|L_i,y_{L_i})  P(L_i,y_{L_i}\vert x',y') {\widetilde{P}}^{(a)}_{z}(x',y') \Big] ,
    \label{eq:absorb_xxx1}
\end{eqnarray}
where we remind readers that the quantity ${\widetilde{P}}_z(x,y|x',y')$ is the $z$-transform of the defect-free propagator.  
On the other hand, the master equation~\eqref{eq:absorb_master_in_t} may be written in the $z$-domain as
\begin{eqnarray}
\fl    {\widetilde{P}}^{(a)}_{z}(x,y) - P^{(a)}_{0}(x,y) = z \sum_{x'}\sum_{y'} \Big[& P(x,y\vert x',y') {\widetilde{P}}^{(a)}_z(x',y')  \nonumber \\
\fl    &- \varrho \sum_{i=1}^{|\mathfrak{S}|} \delta_{x,L_i} \delta_{y,y_{L_i}}   P(L_i,y_{L_i}\vert x',y') {\widetilde{P}}^{(a)}_z(x',y') \Big] ,
\end{eqnarray}
which, for the  values  $x=L_i$ and $y=y_{L_i}$, gives 
\begin{equation}
\fl    {\widetilde{P}}^{(a)}_{z}(L_i,y_{L_i}) - P^{(a)}_{0}(L_i,y_{L_i}) 
    = z(1-\varrho) \sum_{x'} \sum_{y'} P(L_i,y_{L_i}\vert x',y') {\widetilde{P}}^{(a)}_z(x',y') .
\end{equation}
Substituting the above relation in~\eqref{eq:absorb_xxx1}, we obtain
\begin{eqnarray}
\fl    {\widetilde{P}}^{(a)}_{z}(x,y)    &=\sum_{x'}\sum_{y'}  {\widetilde{P}}_z(x,y|x',y') P^{(a)}_0(x',y') \nonumber \\
\fl    & \quad +  \frac{\varrho}{\varrho-1} \sum_{i=1}^{|\mathfrak{S}|} {\widetilde{P}}_{z}(x,y|L_i,y_{L_i})     \Big[ {\widetilde{P}}^{(a)}_{z}(L_i,y_{L_i}) - P^{(a)}_{0}(L_i,y_{L_i}) \Big] .
    \label{eq:absorb_xxx2}
\end{eqnarray}

To (eventually) make connection with the set-up in section~\ref{s:setup}, we now invoke a spatially localized initial condition such that the particle starts from a given site $x_0$ having a state $y_0$ with probability $\alpha_{y_0}$. Therefore, one has~\cite{marris2024persistent}
\begin{equation}
    P^{(a)}_0(x,y) = \delta_{x,x_0} \delta_{y,y_0} \alpha_{y_0} \big[  1-\varrho \, {\mathbf{1}}_{\mathfrak{S}}((x_0,y_0))  \big] ,
    \label{eq:initial0}
\end{equation}
where ${\mathbf{1}}_{\mathfrak{S}}((x_0,y_0))$ is the indicator function defined as ${\mathbf{1}}_{\mathfrak{S}}((x_0,y_0)) = 1$ if $(x_0,y_0) \in {\mathfrak{S}}$, and ${\mathbf{1}}_{\mathfrak{S}}((x_0,y_0)) = 0$ otherwise. Using the initial condition~\eqref{eq:initial0} in~\eqref{eq:absorb_xxx2} and explicitly indicating it in the argument of ${\widetilde{P}}^{(a)}_{z}$, one may obtain
\begin{equation}
\fl    {\widetilde{P}}^{(a)}_{z}(x,y|x_0;y_0)  
    = \alpha_{y_0} {\widetilde{P}}_z(x,y|x_0,y_0)   +  \frac{\varrho}{\varrho-1} \sum_{i=1}^{|\mathfrak{S}|} {\widetilde{P}}_{z}(x,y|L_i,y_{L_i})     {\widetilde{P}}^{(a)}_{z}(L_i,y_{L_i}|x_0;y_0) ,
    \label{eq:absorb_xxx4}
\end{equation}
which is valid for all $x,y$. Here the unusual semi-colon notation indicates that ${\widetilde{P}}^{(a)}_{z}(x,y|x_0;y_0)$ is the $z$-transform of a \emph{weighted} propagator giving the probability for transitioning from an initial site-state pair $(x_0, y_0)$ to $(x,y)$, weighted by the probability $\alpha_{y_0}$ of having that initial state $y_0$. By setting $(x,y)$ in~\eqref{eq:absorb_xxx4} to each of the elements of the set $\mathfrak{S}$ in turn, one obtains a set of four linear coupled equations with four unknown quantities ${\widetilde{P}}^{(a)}_{z}(L_i,y_{L_i}|x_0;y_0)$. 
Solving this set of linear equations via Cramer's rule yields
\begin{eqnarray}
    {\widetilde{P}}^{(a)}_{z}(L,+|x_0;y_0) &=& \alpha_{y_0} (1-\varrho) \frac{\det(\bar{\mathbb{H}}^{(1)}(x_0,y_0,z))}{\det(\bar{\mathbb{H}}(z))}, \nonumber \\
    {\widetilde{P}}^{(a)}_{z}(L,-|x_0;y_0) &=& \alpha_{y_0} (1-\varrho) \frac{\det(\bar{\mathbb{H}}^{(2)}(x_0,y_0,z))}{\det(\bar{\mathbb{H}}(z))} , \nonumber \\[-1ex]
     \label{eq:cramer_sol} \\[-1ex]
    {\widetilde{P}}^{(a)}_{z}(-L,+|x_0;y_0) &=& \alpha_{y_0} (1-\varrho) \frac{\det(\bar{\mathbb{H}}^{(3)}(x_0,y_0,z))}{\det(\bar{\mathbb{H}}(z))}, \nonumber \\
    {\widetilde{P}}^{(a)}_{z}(-L,-|x_0;y_0) &=& \alpha_{y_0} (1-\varrho) \frac{\det(\bar{\mathbb{H}}^{(4)}(x_0,y_0,z))}{\det(\bar{\mathbb{H}}(z))} , \nonumber
\end{eqnarray}
with
{\footnotesize{\begin{eqnarray}
\fl     & \bar{\mathbb{H}}(z) = \nonumber \\
\fl     & \varrho \setlength{\arraycolsep}{1pt}
     \left( \begin{array}{cccc}
     \frac{1-\varrho}{\varrho} + {\widetilde{P}}_{z}(L,+|L,+)  &  {\widetilde{P}}_{z}(L,+|L,-) &  {\widetilde{P}}_{z}(L,+|-L,+) &  {\widetilde{P}}_{z}(L,+|-L,-) \\[1ex]
     {\widetilde{P}}_{z}(L,-|L,+)  & 
     \frac{1-\varrho}{\varrho} +{\widetilde{P}}_{z}(L,-|L,-) &  {\widetilde{P}}_{z}(L,-|-L,+) &  {\widetilde{P}}_{z}(L,-|-L,-) \\[1ex]
     {\widetilde{P}}_{z}(-L,+|L,+) &  {\widetilde{P}}_{z}(-L,+|L,-) & 
     \frac{1-\varrho}{\varrho} +{\widetilde{P}}_{z}(-L,+|-L,+)&  {\widetilde{P}}_{z}(-L,+|-L,-) \\[1ex]
     {\widetilde{P}}_{z}(-L,-|L,+) &  {\widetilde{P}}_{z}(-L,-|L,-) &  {\widetilde{P}}_{z}(-L,-|-L,+)& 
     \frac{1-\varrho}{\varrho} +{\widetilde{P}}_{z}(-L,-|-L,-)
    \end{array} \right) , \nonumber \\
\fl     
\end{eqnarray}}}and $\bar{\mathbb{H}}^{(i)}(x_0,y_0,z)$ being a matrix  with same elements as $\bar{\mathbb{H}}^{(i)}(z)$  except with  its $i$th column replaced by the column vector 
\begin{equation}
\hskip-40pt \Big( {\widetilde{P}}_{z}(L,+|x_0,y_0) \quad {\widetilde{P}}_{z}(L,-|x_0,y_0) \quad {\widetilde{P}}_{z}(-L,+|x_0,y_0) \quad {\widetilde{P}}_{z}(-L,-|x_0,y_0) \Big)^{\rm{T}}  ,
\end{equation}
where the symbol ${\rm{T}}$ stands for matrix transposition. 
As we are interested in perfectly absorbing boundaries, we substitute the solutions~\eqref{eq:cramer_sol} in~\eqref{eq:absorb_xxx4} and then take the limit $\varrho \to 1$ to obtain
\begin{equation}
\fl    {\widetilde{P}}^{(a)}_{z}(x,y|x_0;y_0)  = \alpha_{y_0} {\widetilde{P}}_z(x,y|x_0,y_0)   -  \alpha_{y_0}  \sum_{i=1}^{|\mathfrak{S}|} {\widetilde{P}}_{z}(x,y|L_i,y_{L_i})     \frac{\det(\mathbb{H}^{(i)}(x_0,y_0,z))}{\det(\mathbb{H}(z))} ,
    \label{eq:absorb_xxx5}
\end{equation} 
where $\mathbb{H}(z) = \lim_{\varrho \to 1} \bar{\mathbb{H}}(z)$ and $\mathbb{H}^{(i)}(x_0,y_0,z) = \lim_{\varrho \to 1} \bar{\mathbb{H}}^{(i)}(x_0,y_0,z)$. We emphasize that the right-hand side of~\eqref{eq:absorb_xxx5} has everything written in terms of the known defect-free propagator, as given in~\eqref{eq:Prop_in_xz}. 

The $z$-transform of the absorbing propagator of the run-and-tumble particle starting at $X_0=x_0$ and transitioning to $X_t=x$ irrespective of its hidden states is then obtained by summing over $y$ and $y_0$, i.e., ${\widetilde{P}}^{(a)}_{z}(x|x_0) = \sum_{y} \sum_{y_0}  {\widetilde{P}}^{(a)}_{z}(x,y|x_0;y_0)$. Consequently, the $z$-transform  of the survival probability $Q_{t}(x_0)$ is given by ${\widetilde{Q}}_z(x_0) =\sum_{t=0}^{\infty} z^{t} Q_{t}(x_0)= \sum_{x} {\widetilde{P}}^{(a)}_{z}(x|x_0)$. The (unconditional) FPT probability $\mathsf{P}(\mathcal{T}|x_0)$ for the particle reaching either of the perfectly absorbing sites $L$ or $-L$ starting from $x_0$, irrespective of state, is related to the survival probability by $Q_t(x_0) = 1  - \sum_{\mathcal{T}=0}^{t} \mathsf{P}(\mathcal{T}|x_0) $, where $\mathcal{T}$ is the first-passage time. In the $z$-domain, one may then write $\widetilde{Q}_z(x_0) = 1/(1-z) - {\widetilde{\mathsf{P}}}(z|x_0) / (1-z)$, where ${\widetilde{\mathsf{P}}}(z|x_0)$ is the $z$-transform of the FPT probability. Note that for the FPT we have switched to writing $z$ and $\mathcal{T}$ as arguments, since the time itself is now the random variable of interest rather than a parameter. Using the normalization condition of the defect-free propagator in the $z$-domain, namely, $\sum_{x} \sum_{y} \sum_{y_0} \alpha_{y_0}{\widetilde{P}}_{z}(x,y|x_0,y_0) = \sum_{x} \sum_{y} {\widetilde{P}}_{z}(x,y|x_0) = 1/(1-z)$, one then obtains from~\eqref{eq:absorb_xxx5} that
\begin{equation}
{\widetilde{\mathsf{P}}}(z|x_0) = 1 - (1-z) \widetilde{Q}_z(x_0) = \sum_{y_0} \alpha_{y_0} \sum_{i=1}^{|\mathfrak{S}|}      \frac{\det(\mathbb{H}^{(i)}(x_0,y_0,z))}{\det(\mathbb{H}(z))} .
\label{eq:fpt_gen_with_rho}
\end{equation}
The $z$-transform of the FPT splitting probability to reach the  boundary $L$ and not $-L$, denoted by ${\widetilde{\mathsf{P}}}_+(z|x_0)$, is obtained from~\eqref{eq:fpt_gen_with_rho} by splitting the sum over $i$ to consider only the contribution of the two relevant elements of the set $\mathfrak{S}$. One thus obtains
\begin{eqnarray}
\fl    {\widetilde{\mathsf{P}}}_+(z|x_0)
    = \frac{1}{\det(\mathbb{H}(z))} &  \Big[  \alpha_+ \Big\{     \det(\mathbb{H}^{(1)}(x_0,+,z)) + \det(\mathbb{H}^{(2)}(x_0,+,z))  \Big\} \nonumber \\
\fl    & + \alpha_- \Big\{ \det(\mathbb{H}^{(1)}(x_0,-,z)) + \det(\mathbb{H}^{(2)}(x_0,-,z)) \Big\} \Big] ,  \label{eq:Fp_unnorm}    
\end{eqnarray}
where $\alpha_{+}$ and $\alpha_{-}$ are the probabilities of the particle having the initial states $+$ and $-$, respectively. Similarly, the $z$-transform of the FPT splitting probability to reach the boundary $-L$ and not $L$ is given by
\begin{eqnarray}
\fl    {\widetilde{\mathsf{P}}}_-(z|x_0) = \frac{1}{\det(\mathbb{H}(z))} & \Big[  \alpha_+ \Big\{ \det(\mathbb{H}^{(3)}(x_0,+,z)) + \det(\mathbb{H}^{(4)}(x_0,+,z))  \Big\} \nonumber \\
\fl    & \quad + \alpha_- \Big\{ \det(\mathbb{H}^{(3)}(x_0,-,z)) + \det(\mathbb{H}^{(4)}(x_0,-,z)) \Big\} \Big] . \label{eq:Fm_unnorm}
\end{eqnarray}

The quantities ${\widetilde{\mathsf{P}}}_+(z|x_0)$ and ${\widetilde{\mathsf{P}}}_-(z|x_0)$ are not normalized on their own. However, together they satisfy the normalization condition $\lim_{z \to 1^{-} } \big[ {\widetilde{\mathsf{P}}}_+(z|x_0) + {\widetilde{\mathsf{P}}}_-(z|x_0) \big] = \lim_{z \to 1^{-} } {\widetilde{\mathsf{P}}}(z|x_0) = 1$, which may be verified by taking $z \to 1^{-}$ in~\eqref{eq:fpt_gen_with_rho}. To check the FPT duality, one needs to normalize the individual splitting probabilities by defining 
\begin{equation}
    {\widetilde{P}}_{+}(z|x_0)  = \frac{ {\widetilde{\mathsf{P}}}_+(z|x_0) }{ \lim_{z \to 1^{-} } {\widetilde{\mathsf{P}}}_+(z|x_0) } , \quad  
    {\widetilde{P}}_{-}(z|x_0)  = \frac{ {\widetilde{\mathsf{P}}}_-(z|x_0) }{ \lim_{z \to 1^{-} } {\widetilde{\mathsf{P}}}_-(z|x_0) } ,
    \label{eq:F-z}
\end{equation}
which are the $z$-transforms $\widetilde{P}_\pm(z|x_0) =\sum_{\mathcal{T}=0}^{\infty} z^\T P_\pm(\mathcal{T}|x_0)$ of the FPT probabilities $P_{\pm}(\mathcal{T}|x_0)$ in~\eqref{e:fpt_splitting_def}. For the particular setting of our interest, we fix $x_0=0$, $\alpha_+=p$, and $\alpha_-=q$ and write simply $P_\pm(\mathcal{T})$ for notational brevity.  To obtain $P_\pm(\mathcal{T})$ from $\widetilde{P}_\pm(z)$, one needs to implement an inverse $z$-transform, which, in this case, is a daunting task analytically but can be achieved numerically~\cite{abate_numerical_1992}. 

%================================================
%================================================
%================================================
\section{Numerical results and FPT asymmetry analysis for the RnT model}
\label{s:results}

We now check the analytical expressions obtained in section~\ref{s:fpt-theory} against numerical results obtained from Monte Carlo simulations of large ensembles of trajectories.

The overall drift (mean current) of the RnT particle is easily shown to be independent of $f$ and given by
\begin{equation}
    j=(p-q)(p'-q')=(2p-1)(2p'-1).
\end{equation} For an illuminating comparison we consider throughout this section two parameter sets corresponding to the same drift ($j=0.2$) yet with strikingly different FPT phenomenology, as we show below:
\begin{itemize}
    \item Visible tumbles: parameters $p=0.6$,  $p'=1$ (ballistic runs), and
    \item Hidden tumbles: parameters $p=0.7$,  $p'=0.75$ (diffusive runs)
\end{itemize} 
We choose to label these cases as `visible tumbles' (VT) and `hidden tumbles' (HT) to emphasize the following physical distinction. For VT, every change in the particle's instantaneous velocity is due to a tumble, while for HT a change in the particle's instantaneous velocity may not necessarily be due to a tumble. In other words,  for VT the preferred direction is known from the position trajectory while for HT the preferred direction variable is genuinely hidden; see also the discussion in section~\ref{ss:RnT} where figures~\ref{f:2lane}(b) and~\ref{f:2lane}(c) respectively show example trajectories for VT and HT for $f=0.25$.
    We stress that although the mean currents are the same, the analysis of section~\ref{ss:renewal} predicts asymptotic FPT duality for VT but not HT.

\subsection{Dependence of FPT asymmetry on RnT tumbling probability}
\label{ss:resultsasym}
\begin{figure}
\centering
\includegraphics[scale=0.55]{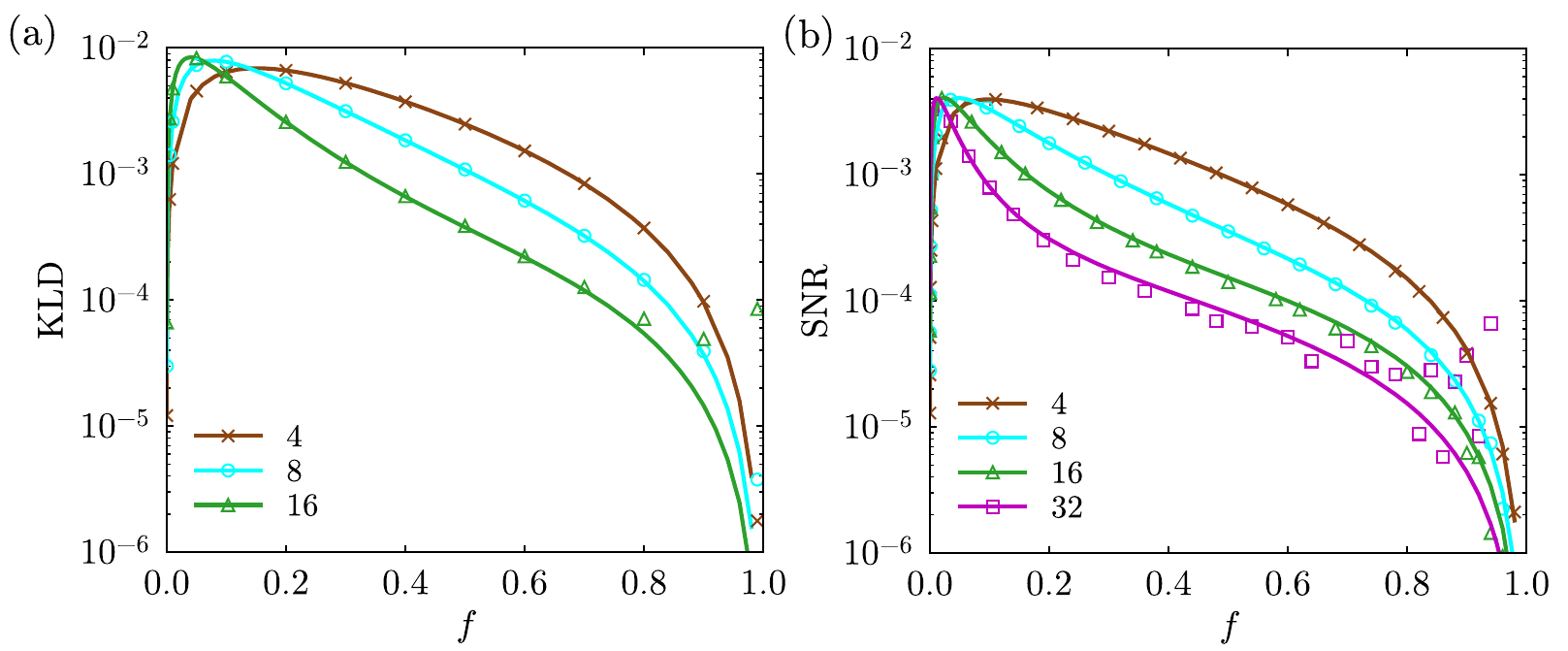}
\caption{FPT asymmetry for the RnT model with visible tumbles (VT, $p=0.6$,  $p'=1$) exiting a symmetric interval. As a function of tumbling probability $f$, we show (a) KLD $D_\mathrm{KL} (P_- || P_+)$ [see~\eqref{eq:kld}] and (b) signal-to-noise ratio $\textrm{SNR}$ [see~\eqref{e:hatrho}] associated with the first-passage times to reach the boundary against and in the direction of the bias.  Exact analytical results are solid lines while estimates from numerical simulations are discrete data points; different colours and symbols represent different boundary positions $L$ (see legend).  Here and in all subsequent figures, simulation data were obtained from $10^9$ trajectories with a cut-off at time 2000 and points only displayed if at least $10^4$ trajectories are absorbed at the negative barrier. The analytical results are obtained by numerically inverting~\cite{abate_numerical_1992} the $z$-transforms of the probabilities $P_{\pm}(\mathcal{T})$ [derived in section~\ref{s:fpt-theory}, see~\eqref{eq:F-z}] and then applying~\eqref{eq:kld} to get the KLD or computing moments to obtain the ${\rm SNR}$ via~\eqref{e:hatrho}.} \label{f:kld} 
\end{figure}

\begin{figure}
\centering
\includegraphics[scale=0.55]{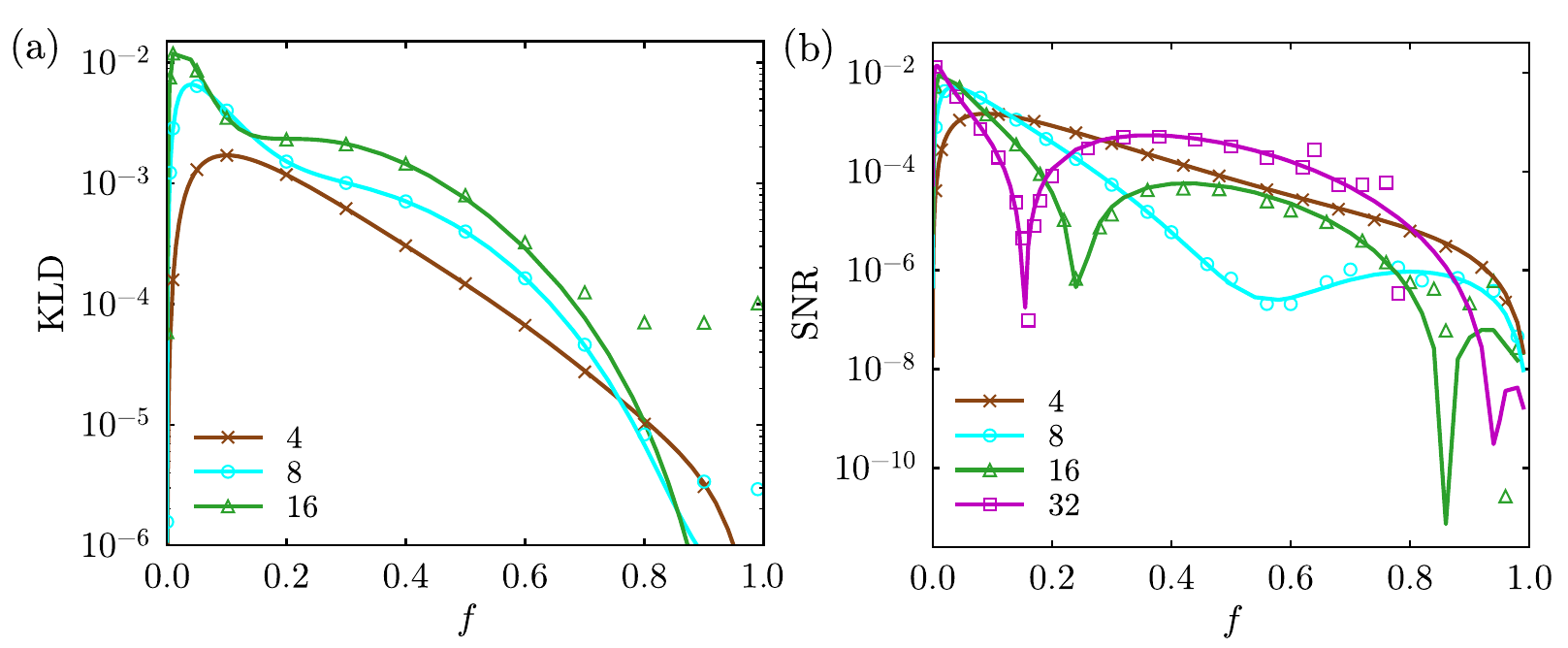}
\caption{FPT asymmetry for the RnT model with hidden tumbles (HT, $p=0.7$,  $p'=0.75$) exiting a symmetric interval. As a function of tumbling probability $f$, we show (a) KLD $D_\mathrm{KL} (P_- || P_+)$ [see~\eqref{eq:kld}] and (b) signal-to-noise ratio $\textrm{SNR}$ [see~\eqref{e:hatrho}] associated with the first-passage times to reach the boundary against and in the direction of the bias.  Exact analytical results are solid lines while estimates from numerical simulations are discrete data points; different colours and symbols represent different boundary positions ($L=4,8,16,32$). The analytical results are obtained by numerically inverting~\cite{abate_numerical_1992} the $z$-transforms of the probabilities $P_{\pm}(\T)$ [derived in section~\ref{s:fpt-theory}, see~\eqref{eq:F-z}] and then applying~\eqref{eq:kld} to get the KLD or computing moments to obtain the ${\rm SNR}$ via~\eqref{e:hatrho}.} \label{f:corr}
\end{figure}

As done often for run-and-tumble models, we  focus our attention here on how quantities of interest vary with the tumbling probability~$f$.  
Specifically, in what follows, we report analytical and numerical results for the
KLD $D_\mathrm{KL} (P_- \vert \vert P_+)$~\eqref{eq:kld}  and the signal-to-noise ratio ${\rm SNR}$~\eqref{e:hatrho}, as a function of $f$ for the VT (figure~\ref{f:kld}) and HT (figure~\ref{f:corr}) scenarios with different systems sizes $L$ (see legend).

For each Monte Carlo trajectory, the simulation was run until the RnT position reached one of the absorbing boundaries or a cut-off time was reached.  The cut-off parameter was chosen very large compared to the mean FPTs so the truncation of distributions should not affect the results.   Indeed we see excellent agreement between the analytical and numerical results except for high $f$ where there is a spurious `upswing' in the numerical results caused, we believe, by poor statistics since here only a small number of trajectories reach the negative boundary.  The fraction of trajectories reaching $-L$ decays exponentially with $L$ so it is challenging to numerically obtain the KLD for systems much larger than those shown here. 

We note that for $f=1$ the model reduces to the ordinary RW (with identical hopping probabilities for the two parameter sets) where duality trivially holds; this is consistent with our observations that $\mathrm{KLD} \to 0$ [see figures~\ref{f:kld}(a) and~\ref{f:corr}(a)] and ${\rm SNR} \to 0$ [see figures~\ref{f:kld}(b) and~\ref{f:corr}(b)] as $f\to 1$ in both scenarios. However, for smaller values of $f$ we find qualitatively different behaviour in the two parameter sets. 

In the VT scenario of figure~\ref{f:kld}, both the KLD and the SNR show a single local maximum (peak) with roughly constant height at a value of $f$ which seems to tend to zero in the large-$L$ limit.  [The scaling of this peak will be examined more closely in the next section.]  Inspection of the underlying data confirms that, within statistical accuracy, $\bar{\tau}_- > \bar{\tau}_+$ for the whole range $0 < f < 1$. For this parameter choice $\bar{\tau}_-$ exceeds $\bar{\tau}_+$ when $p > 0.5$, and vice versa for $p < 0.5$.

On the other hand, in the HT scenario of figure~\ref{f:corr},  the KLD  and the SNR have a more intriguing dependency on the tumbling probability $f$.  
There is still a peak sharpening towards $f=0$ as $L$ increases but there is now evidence of a \emph{second} peak establishing at intermediate $f$ in the large-$L$ limit. For the first peak we have $\bar{\tau}_- > \bar{\tau}_+$ while for the second peak $\bar{\tau}_+ > \bar{\tau}_-$.  The crossover occurs at the point where ${\rm SNR}=0$ which also shifts towards $f=0$ as $L$ increases.

\begin{figure}
\centering
\includegraphics[scale=0.54]{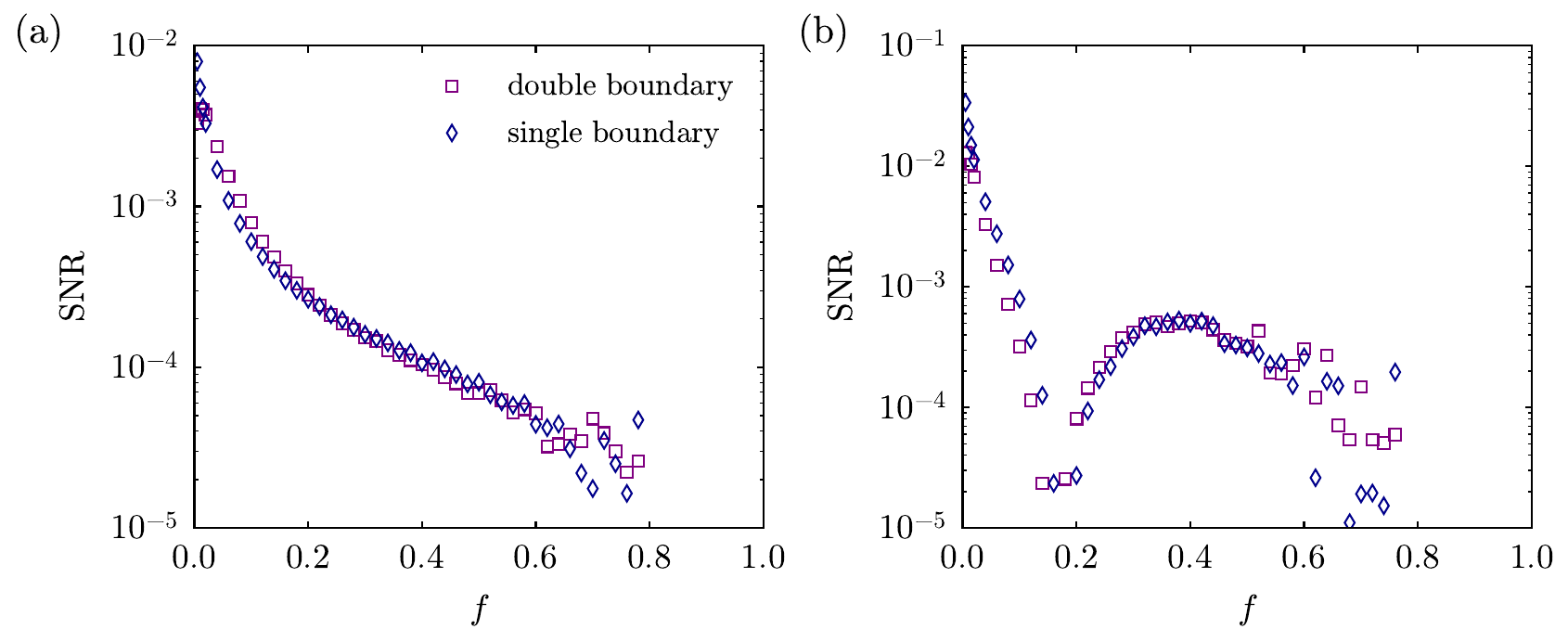}
\caption{Signal-to-noise ratio ${\rm SNR}$ given by~\eqref{e:hatrho} estimated from numerical simulations for the RnT model with parameter sets as in figure~\ref{f:corr}: (a) visible tumbles (VT, $p=0.6$,  $p'=1$), and (b) hidden tumbles (HT, $p=0.7$,  $p'=0.75$).   All points are for the same barrier distance ($L=32$) but showing comparison between double-boundary and single-boundary cases.}
\label{f:singledouble}
\end{figure}

The summary message from figures~\ref{f:kld} and~\ref{f:corr} is that for the HT case we observe the emergence of a second peak for large $L$ whereas it is absent for the VT case: to see this most clearly, the reader is invited to compare the dependence of KLD and SNR on $L$ for intermediate tumbling probabilities (around $f=0.5$) in the two figures.
Significantly, the existence (or otherwise) of this second peak can be predicted by the asymptotic symmetry arguments of section~\ref{ss:renewal}.  In particular, it is easy to check that the SCGF of~\eqref{e:SCGF_RnT} does not satisfy the GC symmetry~\eqref{e:GC} for $p=0.7$, $p'=0.75$ but does for $p=0.6$, $p'=1$. This is consistent with the appearance of the second peak for HT but its absence for VT.  In the VT case, the symmetry will be restored for any finite $f$ as $L \to \infty$, and the first peak concentrates at $f=0$.  [Note that some care is needed with limits here: for any finite $L$, we have $\mathrm{KLD} \to 0$ and ${\rm SNR} \to 0$  as $f \to 0$.]  
In passing, we also remark that our discussion of the relation between GC symmetry and FPT duality included the fact that statistics for double-boundary and single-boundary problems should be asymptotically the same; this is supported by simulation results for $L=32$ in figure~\ref{f:singledouble}.  
  
\subsection{Further comparisons}
\begin{figure}
\centering
\includegraphics[scale=0.55]{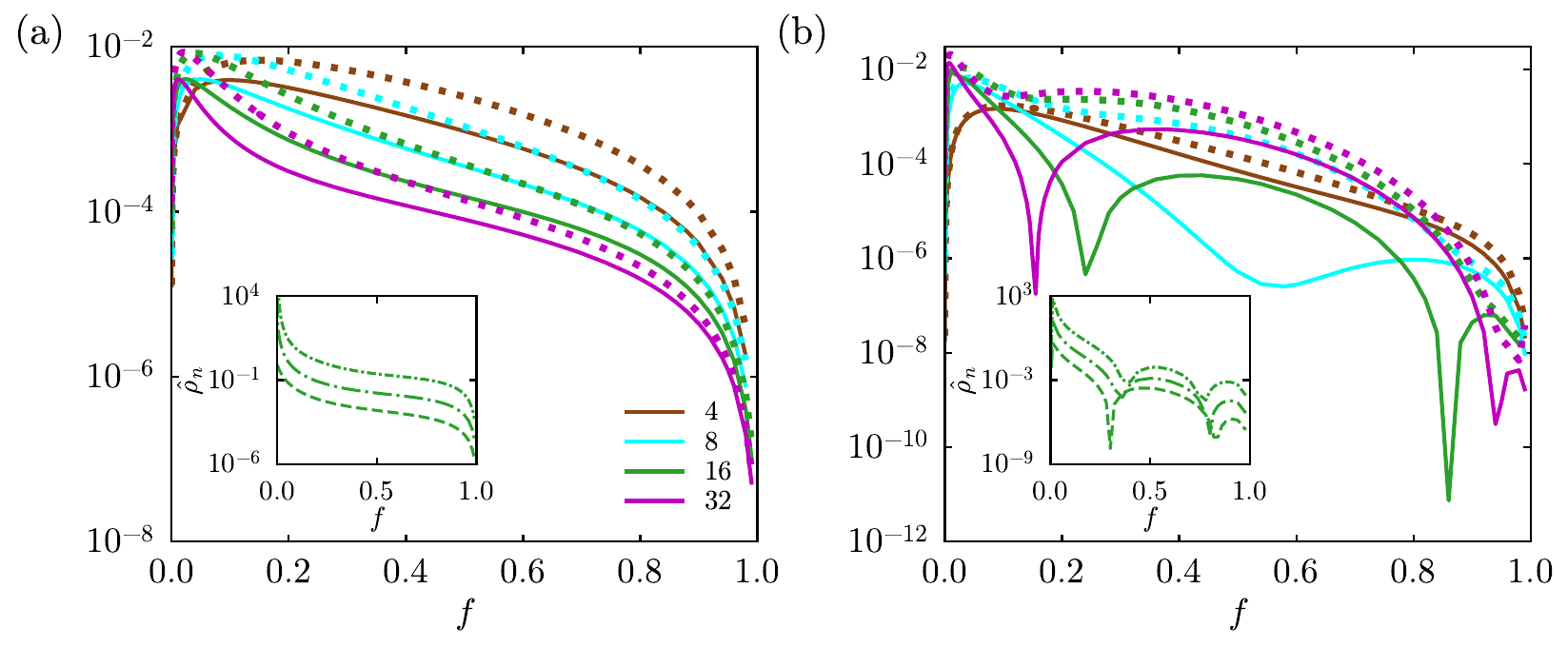}
\caption{Exact analytical value for the KLD $D_\mathrm{KL} (P_- || P_+)$ [dotted lines,~\eqref{eq:kld}] and the signal-to-noise ratio ${\rm SNR}$  [solid lines,~\eqref{e:hatrho}] as a function of the tumbling probability $f$ for the RnT model with two symmetric absorbing boundaries for  (a) visible tumbles (VT, $p=0.6$,  $p'=1$), and (b) hidden tumbles (HT, $p=0.7$,  $p'=0.75$). Different colours represent different barrier positions  ($L=4,8,16,32$). Insets show the scaled squared difference $\hat{\rho}_n$ [see~\eqref{eq:moments_difference}] between the $n$th moments of $P_+(\T)$ and $P_-(\T)$ for $L=16$: dashed lines correspond to $n=2$, dash-dotted lines to $n=3$, and densely dash-dotted lines to $n=4$. The results are obtained by numerically inverting~\cite{abate_numerical_1992} the $z$-transforms of the probabilities $P_{\pm}(\T)$ [derived in section~\ref{s:fpt-theory}, see~\eqref{eq:F-z}] and then applying~\eqref{eq:kld},~\eqref{e:hatrho} and~\eqref{eq:moments_difference}.} 
\label{f:rho_kld}
\end{figure}

The figures in the last subsection suggest that our signal-to-noise ratio ${\rm SNR}$ is indeed a reasonable approximation for the KLD; this is practically important as it is hard to obtain clean simulation data, especially for the KLD.   We emphasize that, in principle, we can obtain results for the RnT with arbitrary interval width via numerically inverting the analytical $z$-transforms~\eqref{eq:F-z} of the probabilities $P_{\pm}(\T)$.  In figure~\ref{f:rho_kld} we use again this approach to more carefully compare the two asymmetry measurements KLD and ${\rm SNR}$ for increasing system sizes.  
The results indicate that the ${\rm SNR}$ provides a lower bound for the KLD throughout the explored parameter range.   In figure~\ref{f:rho_kld}(b) (HT case) we see that the ${\rm SNR}$ appears to have two dips (exaggerated by the log-scale on the vertical axis); one between the `first' and `second' peaks already identified and another for $f$ close to unity which is not seen at all in the KLD.  The positions of these dips move towards zero and one respectively as $L$ increases such that the second peak denominates for large $L$; we will explore this scaling further in the next section. 

It is not surprising that the ${\rm SNR}$ shows slightly different behaviour to the KLD since, as we have highlighted before, the former considers only the first moments of the probabilities $P_{\pm}(\mathcal{T})$ to quantify the asymmetry; in particular the ${\rm SNR}$ will be zero (corresponding to sharp dips on the log-scale plot) whenever the two means are identical even if the distributions are different.  To complement the ${\rm SNR}$ comparison between $P_+(\T)$ and $P_-(\T)$, insets in figure~\ref{f:rho_kld} show the scaled squared difference $\hat{\rho}_n$ [see~\eqref{eq:moments_difference}] between the second, third, and fourth moments, for $L=16$ as a representative example. At least for the first three moments as observed in the numerical results, the magnitude of $\hat{\rho}_n$ generally increases with the order of the moments $n$ in the two parameter sets. For the HT scenario [figure~\ref{f:rho_kld}(b)] we see again evidence of the peak for intermediate~$f$.  

%================================================
%================================================
%================================================
\section{Scaling behaviour}
\label{s:scaling}

In the previous section, we demonstrated that the KLD for the RnT model generically has a non-monotonic form.  In particular, we found two peaks: the first appears to be a finite-size effect (concentrating on $f=0$ as $L \to \infty$) while the second emerges for large system sizes in cases where the FPT duality does not hold, even asymptotically.  These two peaks are also seen in the signal-to-noise ratio ${\rm SNR}$ and, in this section, we examine their scaling more carefully.  We believe there may be connections here to the general FPT bi-scaling theory recently proposed by Baravi et al.~\cite{baravi2023,baravi2025}.

\subsection{First peak}
 
\begin{figure}
\centering
\includegraphics[scale=0.6]{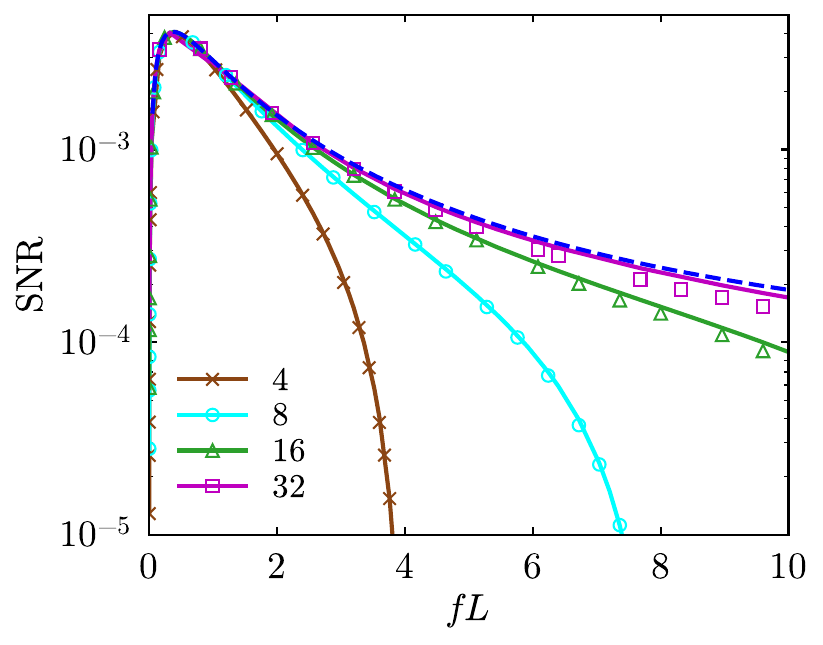}
\caption{Signal-to-noise ratio ${\rm SNR}$ {[see~\eqref{e:hatrho}]} for the RnT model with visible tumbles (VT, $p=0.6$,  $p'=1$); data from figure~\ref{f:kld}(b) but plotted as a function of $fL$. Coloured symbols represent results from numerical simulations for different barrier positions ($L=4,8,16,32$) while solid lines of matching colours are analytical results. The dark blue dashed line is the analytical result for the continuous ballistic run-and-tumble model, see appendix~\ref{s:continuous}. }
\label{f:scale}
\end{figure}

For simplicity we concentrate on the $p'=1$ (visible tumbles) version of the RnT model, where runs are ballistic and there is no second peak.   For fixed $L$, a perturbation analysis for small $f$ (see appendix~\ref{s:perturb}) straightforwardly yields
\begin{equation}
{\rm SNR}\approx\frac{3(p-q)^2fL(L-1)}{4 q(2L-1)},
\end{equation}
which for large $L$ reduces to
\begin{equation}
{\rm SNR}\approx\frac{3(p-q)^2fL}{8 q}. \label{e:rhohatasym}
\end{equation}
The fact that $f$ and $L$ now appear only in the combination $fL$ makes physical sense since this quantity is essentially the probability of a tumble before reaching the boundary and thus might be expected to determine the finite-size behaviour.  Motivated by this fact, together with the observation in section~\ref{ss:resultsasym} that the position of the first peak approaches $f=0$ as $L$ increases, we plot the ${\rm SNR}$ as a function of $fL$ in figure~\ref{f:scale}.  
Notably, we observe a clear collapse of the numerical and analytical data onto a limiting curve as $L\to \infty$.  Furthermore, this limit appears to coincide with the $L$-independent result for an analogous continuous ballistic run-and-tumble model which is also amenable to analytical treatment. The exact moment generating function of the FPT distribution for this continuous model is derived in appendix~\ref{s:continuous}; to the best of our knowledge, this result for asymmetric tumbling rates has not previously been reported in the literature. 

We expect similar phenomenology for the first peak in the $p' \neq 1$ (hidden tumbles) case; figure~\ref{f:corr}(b) supports the assertion that the position of the dip between the two peaks scales (roughly) as $fL$.

\subsection{Second peak}
\label{s:asymptotic}

We return here to examine more closely the large-system behaviour illustrated by the second peak in the RnT model. As previously explained, if the large deviation rate function (or equivalently the SCGF) for the current is known one can immediately determine the presence or absence of the FPT duality in the large-$L$ limit from the presence or absence of the GC symmetry~\eqref{e:GC}.  Furthermore (at least in principle) one can explicitly calculate the asymptotic moments of the conditioned FPTs to either boundary and indeed their large deviations via the correspondence given in~\eqref{e:equiv} and~\eqref{e:equiv2}.  In particular, this then gives a lower bound on the asymptotic KLD, in the form of Kullback's inequality~\eqref{e:Kullback}.

\begin{figure}
\centering
\includegraphics[scale=0.6]{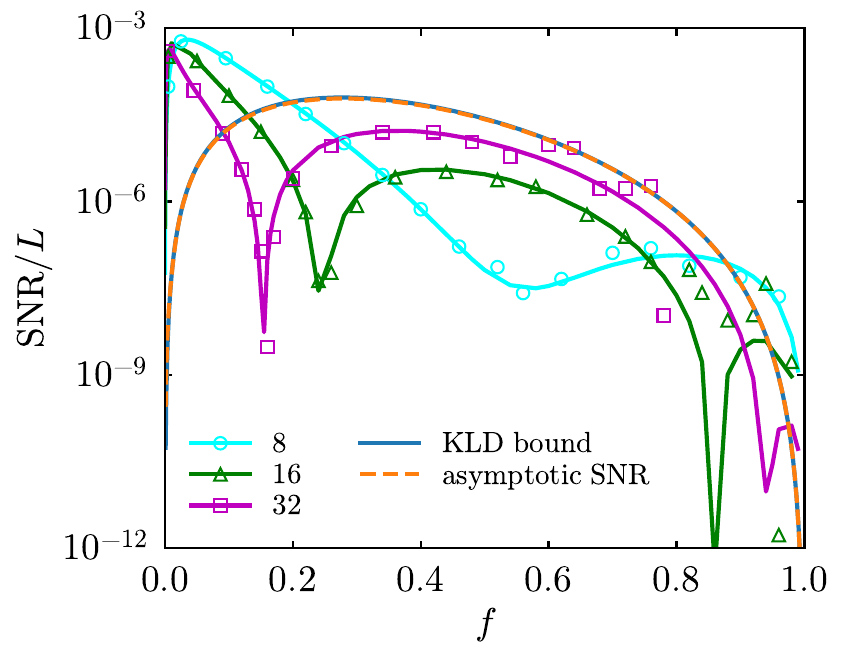}
\caption{Signal-to-noise ratio ${\rm SNR}$ [see~\eqref{e:hatrho}] for the RnT model with hidden tumbles (HT, $p=0.7, p'=0.75$); data from figure~\ref{f:corr}(b) but divided by $L$.  Coloured symbols represent results from numerical simulations for different barrier positions ($L=8,16,32$) while solid lines of matching colours are analytical results.  Solid and dashed lines without symbols are the asymptotic predictions of the left-hand and right-hand sides of~\eqref{e:KLDapprox}, which are effectively coincident at this scale.} 
\label{f:secondpeakscale}
\end{figure}

In figure~\ref{f:secondpeakscale} we test the asymptotic predictions of the previous paragraph for our RnT model, using Mathematica to carry out the numerical Legendre transformation of the known SCGF~\eqref{e:SCGF_RnT} in order to obtain the corresponding rate function. 
To be specific, we show numerical and analytical data for the scaled quantity ${\rm SNR}/L$ in different-sized systems and compare with the prediction calculated from the asymptotic moments [right-hand side of~\eqref{e:KLDapprox}] and with the KLD bound from Kullback's inequality [left-hand side of~\eqref{e:KLDapprox}].  We see that the two sides of~\eqref{e:KLDapprox} are essentially indistinguishable on this scale, presumably because the difference in means is so small that using a Gaussian approximation about the minimum of the rate function does not introduce much error. The data indeed seems to converge towards the asymptotic prediction although accessing larger system sizes would be useful to check this more carefully.

\section{Discussion}
\label{s:conc}

This work extends previous studies on first-passage-time (FPT) duality in Markovian processes to those with a renewal-type structure, focusing particularly on run-and-tumble models as paradigmatic examples of active matter. In this context, we have derived exact analytical formulae for first-passage statistics of a biased discrete run-and-tumble model (RnT) with arbitrary-width intervals, as well as for a continuous analogue.\footnote{Results of this nature also follow from a general connection between exit probabilities and hard walls which is apparently well-known in the mathematical literature but has recently been brought to the attention of physicists in the work of Gu\'neau and Touzo~\cite{Gueneau_2024,Gueneau_2024b}.}  These exact results, supported by Monte Carlo simulations, illustrate two scaling regimes which we believe are generic for such models (compare recent work in~\cite{baravi2023,baravi2025}).  Significantly, the results also reveal that the first-passage properties are not merely determined by the mean current (drift) but strongly affected by the presence of `hidden' variables.  To be concrete, we see that for the studied RnT models on finite intervals, the distribution of first-passage times to the barrier at $+L$ (conditioned on being absorbed at that barrier) is different to the distribution of first-passage times to the barrier at $-L$ (conditioned on being absorbed at that barrier), i.e., there is no FPT duality for finite $L$.   Indeed, for these models, the KLD between conditional distributions shows a `peak' which scales with $fL$ (where $1/f$ is the mean run length) such that it concentrates on $f=0$ as $L\to \infty$. For parameter cases with ballistic runs, meaning that effective tumbles between positive and negative preferred directions are always visible, our analysis suggests that duality is restored as $L \to \infty$ for fixed $f$.  However, for cases with diffusive runs, meaning that effective tumbles between positive and negative preferred directions are genuinely hidden (cannot be inferred from the spatial trajectory), further structure emerges as $L \to \infty$ in the form of a second `peak' in the KLD. Here the FPT duality does not even hold asymptotically. 

The crucial differences in asymptotic behaviour are readily understood with a general approach based on renewal-reward theory which also allows us to extend the analysis to other models. The presence (or absence) of the asymptotic FPT duality for large $L$ is equivalent to the presence (or absence) of the current Galavotti-Cohen (GC) symmetry for large times (as already pointed out for Markovian processes by Gingrich and Horowitz~\cite{gingrich2017fundamental}); in cases where the symmetry does not hold a large-deviation scaling analysis leads to an asymptotic bound on the KLD by Kullback's inequality. Furthermore, at least for processes where the renewal intervals have identical dynamics, and their lengths have finite moments, we have argued that the presence (or absence) of the FPT duality can be determined simply by checking for a particular time-reversal symmetry property in a single renewal interval: the required property is equivalent to the integrated current (position) being proportional to an entropy-like quantity at the end of each renewal interval.  [Technically, the crucial point is the long-time structure of the generating function $G_t(s)$; this object can be simply constructed from knowledge of the generating function in individual renewal intervals.]

\begin{figure}
\centering
\includegraphics[scale=0.6]{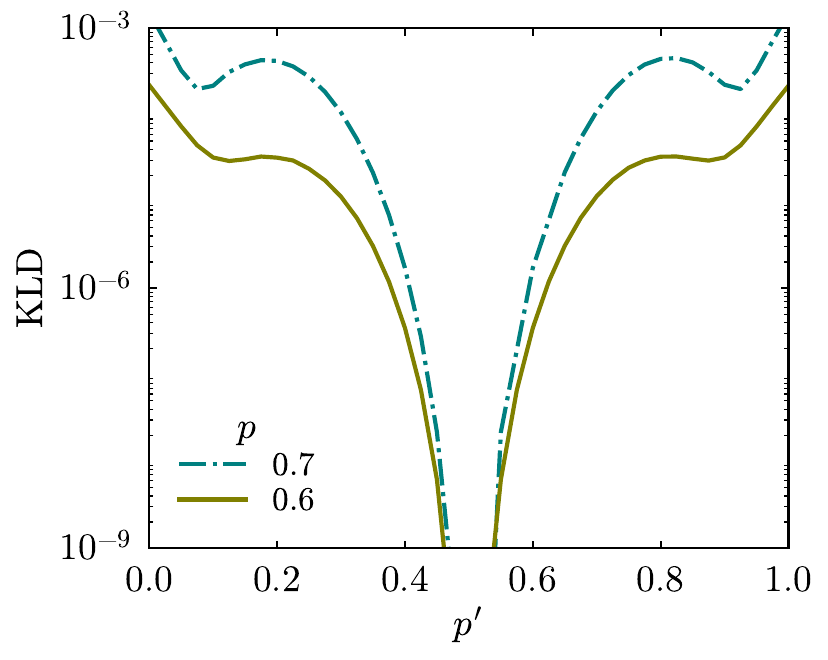}
\caption{Exact analytical value for the  KLD  $D_\mathrm{KL} (P_- || P_+)$ [see~\eqref{eq:kld}] as a function of $p'$ for the RnT model with two symmetric absorbing boundaries, at distance $L=16$, and tumbling probability $f=0.6$.  The results, obtained for two different values of $p$ (see legend), are obtained by numerically inverting~\cite{abate_numerical_1992} the $z$-transforms of the probabilities $P_{\pm}(\mathcal{T})$ [derived in {section}~\ref{s:fpt-theory}, see~\eqref{eq:F-z}] and then applying~\eqref{eq:kld}.}
\label{f:kld_vs_p_prime}
\end{figure}

Although we have primarily focused on the dependence of the KLD on the tumbling probability $f$, our results for the RnT also reveal a non-monotonic dependence on $p'$ (quantifying the bias in each run) for fixed $f$ as shown in  figure~\ref{f:kld_vs_p_prime}. 
For $p'=0.5$, the RnT dynamics is obviously symmetric and therefore the FPT duality is restored.  Away from this point, the non-monotonic dependence on $p'$ implies also a non-monotonic dependence on mean current.

To build up further insight, we now turn to asymptotic predictions for other run-and-tumble models.  In fact, if the tumble mechanism is unaltered from the RnT of section~\ref{ss:RnT} but the dynamics of the runs is changed, then we can obtain the scaled cumulant generating function (SCGF) for the current simply by substituting appropriate forms for $G_\pm(s)$ in~\eqref{e:SCGF_RnT}.
\begin{itemize}
\item An interesting case is an asymmetric model in which the mean current is zero.  Suppose we modify the model of~figure~\ref{f:2lane}(a) such that when the hidden preferred-direction variable is $+$ (i.e., in the upper lane) the particle moves right with probability $p'_+$ and left with probability $q'_+=1-p'_+$, while when the hidden variable is $-$ (i.e., in the lower lane) the particle moves left with probability $p'_-$ and right with probability $q'_-=1-p'_-$.  We now have $G_+(s)=p'_+ e^{-s} + q'_+e^{s}$ and $G_-(s)=p'_- e^{s} + q'_-e^{-s}$.  Choosing $p(p'_+-q'_+)=
q(p'_--q'_-)$ gives zero mean current but it is straightforward to show that, for $p'_+ \neq p'_-$, the current SCGF does not obey the GC symmetry~\eqref{e:GC} and thus there is no FPT duality, even in the large-$L$ limit.  The asymmetry of the underlying hidden dynamics is thus imprinted in the first-passage statistics; the process is non-equilibrium even if the mean current is zero. 

\item For a more obvious asymmetric case, one could take a ballistic model (visible tumbles) with jump length of two units in the positive direction and one unit in the negative direction by setting $G_+(s)=e^{-2s}$ and $G_-(s)=e^s$.  In this case, the asymmetry of jump lengths destroys the GC symmetry; indeed, even if we only consider the position at the end of each renewal interval, the time-reversal of a given trajectory may have probability zero.  Numerical data (not shown) supports the lack of asymptotic FPT duality with evidence for convergence towards the asymptotic prediction from the large deviation argument. 

\item Returning to the case of symmetric (unit) step lengths, one could consider an arguably more biologically-realistic model~\cite{breoni2022one,farago2024confined,kiechl2024transition} in which a latent time is imposed for tumbles.  To be concrete, this means modifying our RnT model such that if a tumble occurs the particle does not move in space (i.e., the diagonal transitions in figure~\ref{f:2lane}(a) become vertical).  For this model the current SCGF was already given in~\cite{shreshtha2019thermodynamic}; interestingly, even for $p'=1$  it does not have the GC symmetry~\eqref{e:GC} and lack of asymptotic FPT duality is again supported by numerical data (not shown).  Here there is also hidden dynamics, in the sense that for two tumbles at consecutive time steps it is impossible to infer the direction chosen in the first tumble from observation of the spatial position; the logarithm of probabilities for a trajectory and its time reversal is not proportional to entropy even if one only considers position at the end of each renewal interval.
\end{itemize}

Physically, the different run-and-tumble variants discussed above indicate that for asymptotic FPT duality, one needs (i) an underlying renewal process where the trajectory of positions at the `beat' times at the end of each renewal interval has non-zero probability under time reversal \emph{and} (ii) for there to be no hidden dynamics in the model.  The first observation is the analogue of the requirement in Markov processes that for GC symmetry~\eqref{e:GC} one needs the reverse transition to be possible for every forward transition.  The second observation is clearly related to the absence of hidden currents required in~\cite{piephoff2025} for the `first-passage time fluctuation theorem' in biomolecular networks.  Hidden currents for the run-and-tumble particle correspond to hidden transitions between $+$ and $-$ states (i.e., hidden effective tumbles).

We stress that for the original RnT model in section~\ref{ss:RnT} of this paper (and the continuous variant in appendix~\ref{s:continuous}, we are able to exactly obtain the corrections to FPT duality for arbitrary interval width.  For more complicated models, such calculations will typically be prohibitive (and there may not be a clear `two-peak' structure in the KLD, since means are not required to cross in general) but nevertheless observation of an asymptotic large-$L$ asymmetry in first-passage times should be regarded as a clear signature of underlying hidden dynamics.   Experimentally, measuring the dependence of asymmetry in first-passage times on interval width $L$, could even be used to infer the timescale of the underlying hidden dynamics (here the run length $1/f$).

Further work is needed to establish general characterization of the finite-size corrections as well as to establish possible connections to other literature on inferring irreversibility in active matter, e.g.,~\cite{PhysRevLett.117.038103, PhysRevX.9.021009, Garcia-Millan_2021,dabelow2025thermodynamicnatureirreversibilityactive}.  More specific open questions relate to dynamics without a renewal structure (e.g., the well-known active Ornstein-Uhlenbeck model~\cite{PhysRevE.103.032607}) as well as to renewal processes where the dynamics depends on the time since renewal or the interval lengths have diverging moments (both scenarios which can lead to dynamical phase transitions~\cite{Harris_2017,Mori_2021}). 
Of course, particle interactions add further complications; the position of a single-tagged particle in a many-particle system (e.g., the zero-range process~\cite{godreche2003dynamics}) is not generally expected to obey the duality although there may be many-particle current-like quantities which do have the symmetry, at least asymptotically.
There is much scope for future studies with both theoretical and practical implications but already our results shed light on the pragmatic yet pioneering approach of using first-passage-time statistics for inference of hidden non-equilibrium properties of active matter~\cite{gladrow2019experimental,hartich2021emergent,kumar2023inference}.
In this contribution, we have restricted our attention to one dimension, but the symmetry and renewal arguments developed here should extend more widely, and it would be valuable to test the robustness of first-passage duality in higher-dimensional active systems.

\ack

This work had its genesis during the London Mathematical Laboratory (LML) Summer School of 2019; we warmly thank Ekaterina Vedenchuk and Jane Garcia for inspiration and preliminary investigations.  We also thank Shamik Gupta for overarching discussions, along with Gunnar Pruessner and Jacob Knight for the particular idea of looking at the asymmetric zero-mean model in section~\ref{s:conc}. 
BW gratefully acknowledges funding from the Imperial College Borland Research Fellowship.  SM was supported by the INSPIRE Fellowship (IF190303) of the Department of Science and Technology, Government of India.
RJH is grateful for further support from the LML in the form of an External Fellowship as well as for hospitality during various stages of the project from the Nordic Institute for Theoretical Physics in Stockholm, the Isaac Newton Institute for Mathematical Sciences in Cambridge (as part of the programme `New statistical physics in living matter: non equilibrium states under adaptive control' supported by EPSRC grant no EP/R014604/1), and the Stellenbosch Institute for Advanced Study. 
ER acknowledges financial support from PNRR MUR project PE00000023-NQSTI. 
\vspace{\baselineskip}

\noindent\emph{Author contributions: ER and RJH conceived the idea and supervised the work; SM made initial numerical observations of the two peaks in the KLD, and did valuable work on their theoretical understanding, while RJH and YS developed later simulations and analysis; RJH carried out the calculations of appendices~\ref{s:PSarg} and~\ref{s:perturb}, as well as contributing the heuristic arguments of the main text; DD carried out the calculations in section~\ref{s:fpt-theory}; BW is responsible for the calculations in appendix~\ref{s:continuous}; DD, RJH, ER, YS, and BW were directly involved in writing the manuscript with further input from SM.}

%================================================
%================================================
%================================================
\begin{appendices}
\section{Renewal argument for current fluctuations in RnT model}
\label{s:PSarg}

In this appendix we present a sketch of the calculations leading to~\eqref{e:SCGF_RnT}, showing in particular how this result emerges within a renewal framework.  The general approach is now well-known for reset dynamics and allows straightforward determination of asymptotic behaviour and identification of dynamical phase transitions (see, e.g.,~\cite{Harris_2017}); indeed it is a direct analogue of the classical Poland-Scheraga (PS) modelling of DNA~\cite{PolandScheraga} with the equilibrium partition function in the PS model now replaced by a non-equilibrium generating function.

\subsection{Underlying method}
\label{ss:PSarg0}

Our aim is to study the moment generating function $G_t(s)= \langle e^{-k X_t} \rangle$ for the total time-integrated current $X_t$ (here just the change in position); the long-time behaviour of this quantity gives the desired SCGF.  We start with the obvious observation that $X_t$ can be decomposed as a sum of the time-integrated currents in each renewal interval [compare~\eqref{e:renewalreward} in the main text]; we can write $X_t = \Delta^{(1)}_{N_1} +\Delta^{(2)}_{N_2}+\Delta^{(3)}_{N_3}+\dots + \Delta^{(F)}_{N_F}$  where $\Delta^{(i)}_{N_i}$ is the time-integrated current in the $i$th renewal interval and we here explicitly indicate the duration $N_i$ of that renewal interval; the number of renewal intervals and their durations are themselves random variables obeying the constraint $t=N_1+N_2+N_3+ \dots + N_F$. [Note that the last renewal interval may be incomplete but if the $N_i$s have finite moments this should not affect the long-time properties; see also the discussion of section~\ref{ss:renewal} in the main text.] 

The basic strategy is to obtain $G_t(s)$ from the weighted generating function for a single renewal interval of known duration: we define $W_n(s)$ as the moment generating function for integrated current in a renewal interval of length $n$ multiplied by the probability of seeing a renewal interval of that length. The desired $G_t(s)$ is then given by products of such weighted generating functions summed over all possible arrangements of intervals which have total duration $t$; in fact, the easiest way to carry out the sum (especially if chiefly interested in the long-time behaviour) is to relax the constraint by switching to Laplace space and defining the $z$-transform of $W_n(s)$ as
\begin{equation}
    \widetilde{W}_z(s) = \sum_{n=1}^\infty z^{n} W_n(s).
\end{equation}
Now the convolutions in the original constrained sum become products and each renewal interval has identical statistics (by definition of the renewal property) so, up to boundary terms, the $z$-transformed moment generating function for $X_t$ is simply
\begin{equation}
\widetilde{G}_z(s)= \widetilde{W}_z(s) + \widetilde{W}_z(s)^2 + \widetilde{W}_z(s)^3 + \widetilde{W}_z(s)^4 + \cdots, \label{e:geo}
\end{equation}
where the first term on the right-hand side corresponds to a trajectory made up of a single renewal interval, the second is a trajectory with two renewal intervals, and so on.  We immediately recognize this as a geometric sum with ratio $\widetilde{W}_z(s)$.  

Taking the inverse $z$-transform we then find
\begin{equation}
    G_t(s) = \frac{1}{2 \pi i} \oint  z^{-t-1} \underbrace{\frac{\tilde{W}_z(s)}{1 - \tilde{W}_z(s)}}_{=\tilde{G}_z(s) } \,\mathrm{d} z \sim [z^*(s)]^{-t}, 
\end{equation}
where we have neglected subleading terms which are exponentially suppressed.
In well-established analogy with the PS model, the SCGF is thus identified with $-\log z^*(s)$ where $z^*(s)$ is the smallest real value of $z$ for which the Laplace transform $\widetilde{G}_z(s)$ diverges.  If there are no dynamical phase transitions, (which could cause relevant divergences in $\widetilde{W}_z(s)$ itself), we see from~\eqref{e:geo} that $z^*(s)$ is just the value of $z$ which solves $\widetilde{W}_z(s)=1$; this turns out to be all we need to consider in the present context.

A remaining subtlety relates to the identification of renewal events which is not unique.  For illustrative purposes we here consider two different decompositions of the RnT into renewal intervals and show that, as expected, they lead to the same result~\eqref{e:SCGF_RnT} for the SCGF.

\subsection{Renewal at all tumbles}
\label{ss:PSarg1}

A natural choice is to consider every tumble event in the RnT as a renewal.   Note that this resets the preferred direction to a distribution (up with probability $p$, down with probability $q$) so the actual dynamics may remain unchanged after the reset.  The inter-renewal times (interval lengths) $N_i$ obviously have the geometric distribution 
\begin{equation}
\mathrm{Prob}(N_i=n)=f(1-f)^{n-1}
\end{equation}
and the weighted moment generating function for the current in a single renewal interval of length $n$ is
\begin{equation}
W^A_n(s)=f(1-f)^{n-1}[pG_+(s)^n+qG_-(s)^n] \label{e:W1}
\end{equation}
where $G_+(s)=p'e^{-s}+q'e^{s}$ is the moment generating function for current in a single time step with preferred direction right (`spin up') while $G_-(s)=p'e^{s}+q'e^{-s}$ is the moment generating function for current in a single time step with preferred direction left (`spin down').   The $z$-transform of~\eqref{e:W1} is
\begin{equation}
    \widetilde{W}^A_z(s)= \frac{fp G_+(s)}{z^{-1}-(1-f)G_+(s)} + \frac{fq G_-(s)}{z^{-1}-(1-f)G_-(s)}
\end{equation}
and setting $\widetilde{W}^A_z(s)=1$ straightforwardly yields a quadratic equation; choosing the positive solution for $z^*(s)$, so that $\mu_\mathrm{RnT}(0)=0$, then gives
\begin{eqnarray}
\fl \mu_\mathrm{RnT}(s) = \log\frac{1}{2}\bigg\{ & (1-f_+)G_-(s)+(1-f_-)G_+(s)  \nonumber \\  
 & + \sqrt{[(1-f_+)G_-(s)+(1-f_-)G_+(s)]^2-4(1-f)G_+(s)G_-(s)}\bigg\}. \label{e:scgfapp}
\end{eqnarray}
with $f_-=fq$ and $f_+=fp$.

\subsection{Renewal at negative-to-positive tumbles}
\label{ss:PSarg2}

An alternative choice which arguably shows the symmetry of the problem more clearly is to split each RnT trajectory at the points where there is an effective tumble from negative to positive.  Significantly, each renewal interval then has identical dynamics: it consists of a portion (length geometrically distributed with parameter $f_-$) with preferred direction right, followed by a portion (length geometrically distributed with parameter $f_+$) with preferred direction left.  At first glance, this makes calculating $W_n(s)$ more involved but again things are easy in the Laplace space where the convolution become a product; we simply have
\begin{equation}
 \widetilde{W}^B_z(s)= \widetilde{U}_z(s)\widetilde{V}_z(s)
\end{equation}
with
\begin{equation}
\fl \widetilde{U}_z(s)= \sum_{n=1}^\infty z^{n} f_-(1-f_-)^{n-1}G_+(s)^n \quad \mathrm{and} \quad \widetilde{V}_z(s)= \sum_{n=1}^\infty z^{n} f_+(1-f_+)^{n-1}G_-(s)^n.
\end{equation}
After carrying out the sums, the result is
\begin{equation}
    \widetilde{W}^B_z(s)=  \frac{f_-G_+(s)}{z^{-1}-(1-f_-)G_+(s)} \times \frac{f_+G_-(s)}{z^{-1}-(1-f_+)G_-(s)}. \label{e:tildeW2}
\end{equation}
Obviously $\widetilde{W}^B_z(s)$ is a different function to $\widetilde{W}^A_z(s)$ but one can readily check that $\widetilde{W}^B_z(s)=1$ reduces to the same quadratic equation as $\widetilde{W}^A_z(s)=1$ and  so the same final expression for $\mu_\mathrm{RnT}(s)$ is recovered as in~\eqref{e:scgfapp}, which is reproduced by~\eqref{e:SCGF_RnT} in the main text.

%================================================
%================================================
%================================================
%================================================
\section{Perturbation argument for RnT model with small tumbling probability}
\label{s:perturb}

In this appendix we examine the small-$f$ behaviour of the signal-to-noise ratio ${\rm SNR}$ in our two-boundary RnT model with $p'=1$ (visible tumbles, ballistic runs).  [A similar analysis could presumably be done for $0 < p' < 1$ (hidden tumbles, diffusive runs) but would be considerably more involved.] Obviously for $f=0$ (and finite $L$) all trajectories reach the boundary without tumbling and we therefore trivially have $\langle T \rangle_- = \langle T \rangle_+$ and ${\rm SNR}=0$.  For fixed $L$ and small $f$ most trajectories will reach the boundary without changing direction (i.e., with no effective tumbles) and we can do a kind of perturbation argument as outlined below.  

We proceed by considering just those trajectories which reach the positive boundary with zero effective tumbles or only one effective tumble; we neglect trajectories with two or more effective tumbles. A no-effective-tumble trajectory which is absorbed at the upper boundary obviously moves only upwards; such a trajectory has probability $p(1-fq)^{L-1}$ and is absorbed after $L$ time steps. A one-effective-tumble trajectory which is absorbed at the upper boundary must have its first step in the negative direction. Recalling that each time step in our model consists of a tumble attempt \emph{followed by} a run move, we see that the single effective tumble can occur at a time given by any integer~$l$ in the interval $[2,L]$; the probability of such a trajectory is $q(1-fp)^{l-2}fp(1-fq)^{L+l-2}$ and the absorption time is $L+2(l-1)$.   Summing over $l$ and normalizing the probabilities we get, in this one-tumble approximation,
\begin{equation}
\fl \langle T \rangle_+ \approx\frac{p(1-fq)^{L-1}  L + \sum_{l=2}^L fpq(1-fp)^{l-2}(1-fq)^{L+l-2}\left[ L+2(l-1)\right]}{p(1-fq)^{L-1} + \sum_{l=2}^L fpq(1-fp)^{l-2}(1-fq)^{L+l-2}} .\label{e:perturb}
\end{equation}

\begin{figure}
\centering
\includegraphics[scale=0.6]{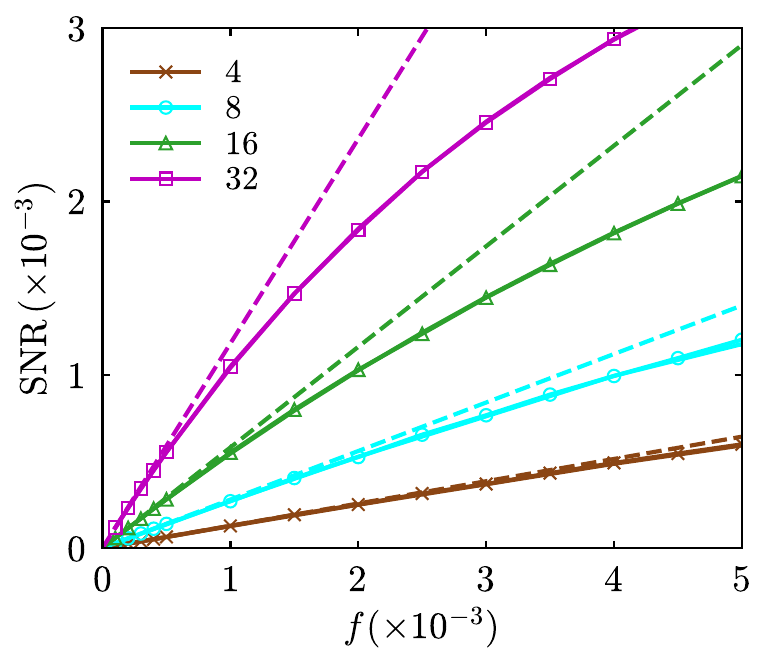}
\caption{Signal-to-noise ratio ${\rm SNR}$  [see~\eqref{e:hatrho}] for the RnT model with visible tumbles (VT, $p=0.6$,  $p'=1$); same data as in figure~\ref{f:kld}(b) but zoomed into the region close to $f=0$.  Coloured symbols represent simulations for different barrier positions ($L=4,8,16,32$) while solid lines of matching colours are analytical results. Dashed lines of matching colours are the analytical small-$f$ predictions of~\eqref{e:rhohat} with the corresponding values of $L$.}
\label{f:perturb}
\end{figure}

Now, dropping terms of order $f^2$ and higher in~\eqref{e:perturb}, we easily conclude
\begin{equation}
\langle T \rangle_+ \approx L + f q L (L-1)
\end{equation}
and, similarly,
\begin{equation}
\langle T \rangle_- \approx L + f p L (L-1).
\end{equation}
An analogous calculation, again to first-order in $f$, yields 
\begin{equation}
\langle T^2 \rangle_+ \approx L^2 + \frac{2}{3} f q L \left(5 L^2 -6 L +1\right)
\end{equation}
and hence
\begin{equation}
\mathrm{Var}_+(T) \approx \frac{2}{3}fq L (2L-1)(L-1).
\end{equation}
Finally, putting everything together, we obtain the leading-order approximation of the ${\rm SNR}$ [as defined in~\eqref{e:hatrho}]:
\begin{equation}
{\rm SNR}\approx\frac{3(p-q)^2fL(L-1)}{4 q(2L-1)}. \label{e:rhohat}
\end{equation}
This approximation is compared with numerical and analytical data in figure~\ref{f:perturb} and indeed predicts the slope of the curve for small $f$. 

Note that for the equivalent single-boundary case, the upper limit in the sums of~\eqref{e:perturb} would be infinity and in fact these sums would diverge; a trajectory can make an arbitrarily long excursion in the opposite direction before tumbling.  This suggests that ${\rm SNR} \to \infty$ as $f \to 0$ in this case, consistent with preliminary simulation data, see e.g., figure~\ref{f:singledouble}.

%================================================
%================================================
%================================================
\section{Derivation of exact FPT statistics for a continuous run-and-tumble variant with two absorbing boundaries}
\label{s:continuous}

In this appendix, we analyse a variant run-and-tumble model in continuous time and continuous space to show that our general approach is valid beyond the discrete setting which is the focus of the main text. As usual, we consider the exit problem of the particle in a two-sided interval $[-L, L]$.   We take asymmetric ballistic run-and-tumble motion (`visible tumbles') for which the FPT duality is expected to hold in the wide-interval limit, and demonstrate that one can obtain the FPT statistics exactly to quantify the finite-size violations.  The recent paper of Neri~\cite{Neri_2025} also considered a continuous run-and-tumble model but with \emph{symmetric} dynamics and asymmetric boundaries.  We do the analysis here for an arbitrary initial position in the interval (implicitly therefore also including asymmetric boundaries) but for contact with the set-up of section~\ref{s:setup} in the main text we will ultimately be interested in FPT statistics starting from the origin.

\subsection{Asymmetric ballistic run-and-tumble motion}
We consider an asymmetric continuous version of the paradigmatic run-and-tumble process \cite{Berg_2004}: a particle moves ballistically in one dimension while its direction undergoes instantaneous tumbling at randomly spaced points in time. A straightforward way to implement a drift in the particle's average position is to choose different speeds for different directions as discussed for instance in~\cite{debruyne2021c}; in that case, however, each trajectory has zero measure under spatial path reflection leading, for the FPT distributions, to a diverging KLD $D_\mathrm{KL} (P_- || P_+)$ [see~\eqref{eq:kld}], for arbitrarily small biases. In order to circumvent this, we instead construct a run-and-tumble process with biased expected position where the asymmetry stems from unequal tumbling rates. Such a situation is analogous to having $p \neq q$ in the discrete RnT of the main text.

Formally, we write the continuous model as
\begin{equation}
\label{eq:asymm_rnt}
    \dot{X}_t = v Y_t,
\end{equation}
where $Y_t \in \left \lbrace +, - \right \rbrace$ is a telegraphic noise~\cite{gardiner2009stochastic,Basu2024Target} tumbling from the state $-$ ($+$) to the state $+$ ($-$) with rate $\gamma^+$ ($\gamma^-$). Note that we consider the generic case in which $\gamma^+$ and $\gamma^-$ are both strictly positive and may be either equal or different to each other. We further introduce the average tumbling rate $\gamma$ and rate bias $\Gamma$:
\begin{equation}
\label{appeq:def_gamma_delta}
    \gamma = \frac{\gamma^+ + \gamma^-}{2}, \qquad 
    \Gamma = \gamma^+ - \gamma^-,
\end{equation}
which, by construction, satisfy $|\Gamma| < 2 \gamma$. At stationarity, the probability to find the particle in a positive (negative) state is given by
\begin{equation}
\label{eq:steady_state_probs}
    p_{+} = \frac{\gamma^+}{\gamma^+ + \gamma_-} = \frac{1}{2} + \frac{\Gamma}{4 \gamma} ,
    \qquad
    p_{-} = \frac{\gamma^-}{\gamma^+ + \gamma_-} = \frac{1}{2} - \frac{\Gamma}{4 \gamma} .
\end{equation}

\subsection{Laplace transforms of the first-passage density in a two-sided interval}

We here initialize the particle at a general inner position in $(-L, L)$ and,   
following standard methods (see, e.g., \cite{malakarSteadyStateRelaxation2018a}), separately consider the FPT density conditioned on (i) the starting internal state ($+$ or $-$) and (ii) the boundary ($+L$ or $-L$) from which it leaves the interval. Specifically, we denote by $\mathsf{P}_{\pm}(\mathcal{T}|x,y)$ the probability density that the particle, starting from position $x$ with internal state $y$, first reaches the boundary at $\pm L$ a time $\mathcal{T}$ later.

We set out by constructing the backward master equations for these four FPT densities taking advantage of the renewal property: 
\begin{eqnarray}
\fl \mathsf{P}_+(\mathcal{T}|x,+) &= (1-\gamma^-  \dd{\mathcal{T}})\mathsf{P}_+(\mathcal{T}-\dd{\mathcal{T}}|x+v \dd{\mathcal{T}},+) + \gamma^-  \dd{\mathcal{T}} \mathsf{P}_+(\mathcal{T}-\dd{\mathcal{T}}|x,-) , \\
\fl \mathsf{P}_+(\mathcal{T}|x,-) &= (1-\gamma^+  \dd{\mathcal{T}})\mathsf{P}_+(\mathcal{T} -\dd{\mathcal{T}}|x-v \dd{\mathcal{T}},-) + \gamma^+  \dd{\mathcal{T}} \mathsf{P}_+(\mathcal{T}-\dd{\mathcal{T}}|x,+) ,\\
\fl \mathsf{P}_-(\mathcal{T}|x,+) &= (1-\gamma^-  \dd{\mathcal{T}})\mathsf{P}_-(\mathcal{T}-\dd{\mathcal{T}}|x+v \dd{\mathcal{T}},+) + \gamma^-  \dd{\mathcal{T}} \mathsf{P}_-(\mathcal{T}-\dd{\mathcal{T}}|x,-) ,  \\
\fl \mathsf{P}_-(\mathcal{T}|x,-) &= (1-\gamma^+  \dd{\mathcal{T}})\mathsf{P}_-(\mathcal{T}-\dd{\mathcal{T}}|x-v \dd{\mathcal{T}},-) + \gamma^+  \dd{\mathcal{T}} \mathsf{P}_-(\mathcal{T}-\dd{\mathcal{T}}|x,+) .
\end{eqnarray}
Collecting expansion terms of first order in $\dd{\mathcal{T}}$, one obtains that the densities satisfy the first-order partial differential equations
\begin{eqnarray}
\label{eq:pde_H1}
     \partial_\mathcal{T} \mathsf{P}_{\pm}(\mathcal{T}|x,+) &= -\gamma^- \mathsf{P}_{\pm}(\mathcal{T}|x,+) + \gamma^- \mathsf{P}_{\pm}(\mathcal{T}|x,-) + v \partial_x \mathsf{P}_{\pm}(\mathcal{T}|x,+) ,\\
\label{eq:pde_H2}
     \partial_\mathcal{T} \mathsf{P}_{\pm}(\mathcal{T}|x,-) &= -\gamma^+ \mathsf{P}_{\pm}(\mathcal{T}|x,-) + \gamma^+ \mathsf{P}_{\pm}(\mathcal{T}|x,+) - v \partial_x \mathsf{P}_{\pm}(\mathcal{T}|x,-),
\end{eqnarray}
together with the boundary conditions
\begin{eqnarray}
\label{eq:h_init_1}
    &\mathsf{P}_+(\mathcal{T}|L,+) = \delta(\mathcal{T}) , \qquad
    &\mathsf{P}_-(\mathcal{T}|-L,-) = \delta(\mathcal{T}) ,\\
\label{eq:h_init_2}
    &\mathsf{P}_+(\mathcal{T}|-L,-) = 0  , \qquad
    &\mathsf{P}_-(\mathcal{T}|L,+) = 0,
\end{eqnarray}
and the initial conditions $\mathsf{P}_{\pm}(0|x,\pm) 
 = \mathsf{P}_{\pm}(0|x,\mp) = 0$ for $x \in (-L,L)$.

Next, we apply a Laplace transform in time, defined by $\widetilde{f}(s|x,y) = \int_0^{\infty} e^{-s\mathcal{T}} f(\mathcal{T}|x,y) \,\dd{\mathcal{T}}$, to~\eqref{eq:pde_H1} and \eqref{eq:pde_H2}.\footnote{Recall that the Laplace transform of a probability density function gives the moment generating function; it is analogous to the $z$-transform in discrete time, with $e^{-s}$ mapped to $z$, so we use the same $\widetilde{\mathsf{P}}$ notation here.} This yields, for $x\in (-L,L)$,
\begin{eqnarray}
\label{eq:h_matching_1}
    \left( s + \gamma^- - v \partial_x \right)\widetilde{\mathsf{P}}_\pm(s | x,+) &= \gamma^- \widetilde{\mathsf{P}}_\pm(s|x,-) ,\\
\label{eq:h_matching_2}
    \left( s + \gamma^+ + v \partial_x \right)\widetilde{\mathsf{P}}_\pm(s|x,-) &= \gamma^+ \widetilde{\mathsf{P}}_\pm(s|x,+).
\end{eqnarray}
Combining the corresponding sets of coupled first-order ordinary differential equations, from~\eqref{eq:h_matching_1} and~\eqref{eq:h_matching_2}, one may obtain
\begin{eqnarray}
\left[ \left( s + \gamma^+ + v \partial_x \right) \left( s + \gamma^- - v \partial_x \right) - \gamma^+ \gamma^-\right] \widetilde{\mathsf{P}}_{\pm}(s|x,+) &= 0 ,\\
 \left[ \left( s + \gamma^- - v \partial_x \right) \left( s + \gamma^+ + v \partial_x \right) - \gamma^+ \gamma^- \right] \widetilde{\mathsf{P}}_{\pm}(s|x,-) &= 0 ,
\end{eqnarray}
which give the same second-order ordinary differential equation for all four transformed densities. Using~\eqref{appeq:def_gamma_delta}, we get 
\begin{eqnarray}
\label{eq:2nd-order-decoupled}
  \left[ - v^2 \partial_x^2 - \Gamma v \partial_x + s^2 + 2 \gamma s  \right] \widetilde{\mathsf{P}}_{\pm}(s|x,y) &= 0 ,
\end{eqnarray}
with $y \in \{+,-\}$.
In order to solve~\eqref{eq:2nd-order-decoupled}, one chooses an exponential ansatz $\widetilde{\mathsf{P}}_\pm(s|x,y) \propto\exp(\lambda(s) x)$ and finds the admissible choices of $\lambda$ as
\begin{eqnarray}
\label{eq:lambda_pm}
    \lambda^{\pm}(s) = - \frac{1}{v} \left[\frac{\Gamma}{2} \pm \sqrt{s^2 + 2 \gamma s + \frac{\Gamma^2}{4}} \right]  .
\end{eqnarray}

The desired solutions for the four transformed FPT densities are then given by the linear combinations
\begin{eqnarray}
    \widetilde{\mathsf{P}}_+(s|x,y=+) &= A_1 \exp ( \lambda^+ x ) + A_2 \exp ( \lambda^- x ) ,\\
    \widetilde{\mathsf{P}}_+(s|x,y=-) &= B_1 \exp ( \lambda^+ x ) + B_2 \exp ( \lambda^- x ) ,\\
    \widetilde{\mathsf{P}}_-(s|x,y=+) &= C_1 \exp ( \lambda^+ x ) + C_2 \exp ( \lambda^- x ) ,\\
    \widetilde{\mathsf{P}}_-(s|x,y=-) &= D_1 \exp ( \lambda^+ x ) + D_2 \exp ( \lambda^- x )  . 
\end{eqnarray}
The eight coefficients are determined by the four boundary conditions of \eqref{eq:h_init_1} and \eqref{eq:h_init_2}, which imply (after Laplace transform)
\begin{eqnarray}
\label{eq:coeff-eq1}
    A_1 \exp\left( \lambda^+ L \right) + A_2 \exp\left( \lambda^- L \right)= 1 ,\\
    B_1 \exp \left( - \lambda^+ L \right) + B_2 \exp \left( - \lambda^- L \right)
    = 0 , \\
    C_1 \exp \left( \lambda^+ L \right) + C_2 \exp \left( \lambda^- L \right) 
    = 0 ,\\
    D_1 \exp \left( - \lambda^+ L \right) + D_2 \exp \left(- \lambda^- L \right) 
= 1  ,
\end{eqnarray}
as well as the matching conditions of~\eqref{eq:h_matching_1} and~\eqref{eq:h_matching_2}, which lead to
\begin{eqnarray}
    A_1 &= \frac{\left(\gamma - \frac{\Gamma}{2 } \right)}{s+\left(\gamma - \frac{\Gamma}{2 } \right) - v \lambda^+} B_1 ,\\
    A_2 &= \frac{\left(\gamma - \frac{\Gamma}{2 } \right)}{s+\left(\gamma - \frac{\Gamma}{2 } \right) - v \lambda^-} B_2 ,\\
    D_1 &= \frac{\left(\gamma + \frac{\Gamma}{2} \right)}{s + \left(\gamma + \frac{\Gamma}{2} \right) + v \lambda^+} C_1 ,\\
    D_2 &= \frac{\left(\gamma + \frac{\Gamma}{2} \right)}{s + \left(\gamma + \frac{\Gamma}{2} \right) + v \lambda^-} C_2 . \label{eq:coeff-eq8}
\end{eqnarray}
Although cumbersome, the eight independent equations~\eqref{eq:coeff-eq1}--\eqref{eq:coeff-eq8} can be solved using standard computer algebra systems. As the expressions themselves are quite lengthy, we do not report them here. 

It remains to determine the (Laplace-transformed) first-passage densities to the positive boundary $+L$ and the negative boundary $-L$ conditioned on the particle being initialized at position $x_0$ in the steady state. The result is given by
\begin{eqnarray}
    \widetilde{\mathsf{P}}_\pm(s|x_0) = p_+ \widetilde{\mathsf{P}}_\pm(s|x_0,+) + p_- \widetilde{\mathsf{P}}_\pm(s|x_0,-)  ,
\end{eqnarray}
where the quantities $p_{\pm}$ are given in~\eqref{eq:steady_state_probs}. The FPT density to reach the positive boundary is then obtained as
\begin{eqnarray}
\label{appeq:h_upper}
\fl  \widetilde{\mathsf{P}}_+(s|x_0) =  e^{ \frac{\Gamma}{2v}\left( L-x_0\right)} \left(\frac12 + \frac{\Gamma}{4 \gamma} \right)  \frac{\phi(s) \cosh \left[\frac{L+x_0}{v }\phi(s) \right] +  \left(s + 2\gamma - \frac{\Gamma}{2}\right) \sinh\left[ \frac{L+x_0}{v }\phi(s)  \right]}{\phi(s) \cosh \left[\frac{2L}{v } \phi(s) \right] + (s + \gamma)\sinh \left[\frac{2L}{v } \phi(s) \right]  } , \nonumber \\
\end{eqnarray}
where we have defined
\begin{equation}
    \phi(s) = \sqrt{s^2 + 2\gamma s + \frac{\Gamma^2}{4}} .
\end{equation}
The FPT density to reach the negative boundary $\widetilde{\mathsf{P}}_-(s|x_0)$ conditioned on the particle being
initialized at position $x_0$ in the steady state then follows from the fundamental reflection symmetry
  \begin{equation}
  \label{appeq:reflection_symmetry}
  \widetilde{\mathsf{P}}_-^{(\Gamma)}(s|x_0) =  \widetilde{\mathsf{P}}_+^{(-\Gamma)}(s|-x_0), 
\end{equation}
under simultaneous reflections of $x_0 \to -x_0$ and $\Gamma \to - \Gamma$. 

\subsection{Conditional first-passage times}
The Laplace-transformed functions $\widetilde{\mathsf{P}}_{+}(s|x_0)$ and $\widetilde{\mathsf{P}}_-(s|x_0)$ provide the unnormalized moment generating functions of the FPT densities to reach the positive and the negative boundaries, respectively. Conditioning on the boundary through which the particle leaves the interval, one may define the normalized \emph{conditional} moment generating functions 
\begin{equation}
\label{appeq:def_H_plus}
    \widetilde{P}_+(s|x_0) = \frac{\widetilde{\mathsf{P}}_+(s|x_0)}{\lim_{s \to 0^+} \widetilde{\mathsf{P}}_+(s|x_0) } , \qquad 
    \widetilde{P}_-(s|x_0) = \frac{\widetilde{\mathsf{P}}_-(s|x_0)}{\lim_{s \to 0^+} \widetilde{\mathsf{P}}_-(s|x_0) }  \ .
\end{equation}
We remark that~\eqref{appeq:def_H_plus} is completely analogous to~\eqref{eq:F-z} for the discrete spatiotemporal run-and-tumble dynamics analysed in section~\ref{s:fpt-theory}. The symmetry outlined in~\eqref{appeq:reflection_symmetry} also holds here: $  \widetilde{P}_-^{(\Gamma)}(s|x_0) =  \widetilde{P}_+^{(-\Gamma)}(s|-x_0)$. Without loss of generality, it hence suffices to study
\begin{eqnarray}
\label{appeq:Mplus_expression}
    \widetilde{P}_+(s|x_0) &=   \frac{ |\Gamma| \cosh\left(\frac{L}{v} |\Gamma| \right) + 2 \gamma \sinh \left(\frac{L}{v} |\Gamma| \right)}{|\Gamma| \cosh \left(\frac{L+x_0}{2v} |\Gamma| \right) + (4 \gamma - \Gamma) \sinh\left(\frac{L+x_0}{2v} |\Gamma| \right)} \nonumber \\
    &\hskip5pt  \times \frac{\phi(s) \cosh \left[\frac{L+x_0}{v }\phi(s) \right] +  \left(s + 2\gamma - \frac{\Gamma}{2}\right) \sinh\left[ \frac{L+x_0}{v }\phi(s)  \right]}{\phi(s) \cosh \left[\frac{2L}{v } \phi(s) \right] + (s + \gamma)\sinh \left[\frac{2L}{v } \phi(s) \right]  },
\end{eqnarray}
which follows from~\eqref{appeq:h_upper} and~\eqref{appeq:def_H_plus}.

The asymmetry of the conditional moment generating functions under drift-reversal is now revealed in the explicit dependence of~\eqref{appeq:Mplus_expression} on the sign of $\Gamma$.  [This is in contrast to what would be found doing the same calculation for simple biased diffusion.]  To make contact with the main text, we can set $x_0=0$ and quantify the asymmetry by calculating the signal-to-noise ratio ${\rm SNR}$ as in section~\ref{s:asymmetry}. To this end, we note that the conditional moment generating functions produce the \emph{conditional} $n$th moment of the FPT to reach the positive and the negative boundaries, given by
\begin{equation}
\label{appeq:first_conditional_moment}
    \expval{T^n}_+ = (-1)^n \dv{n}{}{s} \widetilde{P}_+(s) \Big|_{s=0}  \quad  \mathrm{and} \quad
    \expval{T^n}_- = (-1)^n \dv{n}{}{s} \widetilde{P}_-(s) \Big|_{s=0},
\end{equation}
respectively. The required expressions for the first two conditional moments are rather unwieldy and not reported here, but can easily be derived from~\eqref{appeq:Mplus_expression} and~\eqref{appeq:first_conditional_moment}. This approach was used to obtain the dashed line for the ${\rm SNR}$ in figure~\ref{f:scale} of the main text, setting $\gamma=f/2$ and $\Gamma=f(p-q)$ to match parameters.  In principle, of course, one could also calculate the full KLD by (numerically) inverting the Laplace-transformed densities.

By inspection of $\widetilde{P}_\pm(s)$, one can also directly obtain information about the dependence of the asymmetry on system size.  Firstly, note that $\widetilde{P}_\pm(s/L)$ gives the Laplace transform of the conditional distributions of $T/L$, i.e., the \emph{scaled} FPT.  Secondly, note that if we also make the substitutions $\gamma \to \gamma/L$ and $\Gamma \to \Gamma/L$ in $\widetilde{P}_\pm(s/L)$, the expressions become $L$-independent (for $x_0=0$).  This explains the fact that the curve of ${\rm SNR}$ against $fL$ in figure~\ref{f:scale} is $L$-independent.  If the same curve was plotted against $f$ (or $\gamma$), the peak would concentrate around zero as $L \to \infty$ consistent again with the asymptotic restoration of duality. Indeed, one may verify  that $\lim_{L\to \infty} \widetilde{P}_+(s/L) = \lim_{L\to \infty} \widetilde{P}_-(s/L) = \exp(-2 s \gamma/ v |\Gamma|)$.

\end{appendices}

\section*{References}
\bibliographystyle{iopart-num} 
\bibliography{refs}

\end{document}